\documentclass[acmsmall,authorversion,nonacm]{acmart}

\AtBeginDocument{%
	}

\setcopyright{acmcopyright}
\copyrightyear{2022}
\acmYear{2022}
\acmDOI{XXXXXXX.XXXXXXX}

\acmJournal{JACM}

\usepackage{float}

\usepackage{subfig}
\usepackage{enumitem}

\usepackage{booktabs} 

\usepackage{csquotes}
\usepackage{amsmath}
\usepackage{outlines}

\usepackage{siunitx}
\def\capfirstletteraux#1#2\relax{\uppercase{#1}\lowercase{#2}}

\usepackage{marvosym}

\usepackage{graphicx}
\usepackage{graphbox} 

\usepackage{array}
\newcolumntype{P}[1]{>{\centering\arraybackslash}m{#1}}
\newcolumntype{L}[1]{>{\arraybackslash}m{#1}}
\newcolumntype{R}{>{\raggedleft\arraybackslash}X}

\usepackage{xspace}
\newcommand\ie{i.\,e.\xspace}
\newcommand\eg{e.\,g.\xspace}

\newcommand\US{U.\,S.\xspace}

\newcommand{\mathup}[1]{\mathrm{#1}}

\usepackage{url}
\usepackage[normalem]{ulem}
\usepackage{balance} 

\makeatletter
\renewcommand{\fps@figure}{htb}         
\renewcommand{\fps@table}{htb}         
\makeatother 

\newcommand\mc[1]{\multicolumn{1}{c}{#1}}

\usepackage{tabularx}

\allowdisplaybreaks

\newcommand{\citeeg}[1]{(e.g., \citeauthor{#1} \citeyear{#1})}

\begin{document}

\title{Diffusion of Community Fact-Checked Misinformation on Twitter}


\author{Chiara Drolsbach}
\affiliation{%
	\institution{JLU Giessen}
	\country{Germany}}
\email{chiara.drolsbach@wi.jlug.de}

\author{Nicolas Pröllochs}
\affiliation{%
	\institution{JLU Giessen}
	\country{Germany}}
\email{nicolas.proellochs@wi.jlug.de}


\begin{abstract}
The spread of misinformation on social media is a pressing societal problem that platforms, policymakers, and researchers continue to grapple with. As a countermeasure, recent works have proposed to employ non-expert fact-checkers in the crowd to fact-check social media content. While experimental studies suggest that crowds might be {able} to accurately assess the veracity of social media content, an understanding of how crowd fact-checked (mis-)information {spreads} is missing. In this work, we empirically analyze the spread of misleading vs. not misleading community fact-checked posts on social media. For this purpose, we employ a dataset of community-created fact-checks from Twitter's ``Birdwatch'' pilot and map them to resharing cascades on Twitter. Different from earlier studies analyzing the spread of misinformation listed on third-party fact-checking websites (\eg, snopes.com), we find that community fact-checked misinformation is {less} viral. Specifically, misleading posts are estimated to receive \SI{36.62}{\percent} fewer retweets than not misleading posts. A partial explanation may lie in differences in the fact-checking targets: community fact-checkers tend to fact-check posts from influential user accounts with many followers, while expert fact-checks tend to target posts that are shared by less influential users. We further find that there are significant differences in virality across different sub-types of misinformation (\eg, factual errors, missing context, manipulated media). Moreover, we conduct a user study to assess the perceived reliability of (real-world) community-created fact-checks. Here, we find that users, to a large extent, agree with community-created fact-checks. Altogether, our findings offer insights into how misleading vs. not misleading posts spread and highlight the crucial role of sample selection when studying misinformation on social media. 
\end{abstract}

%



\maketitle

\section{Introduction}


There are widespread concerns that misinformation on social media is damaging societies and democratic institutions \cite{Lazer.2018}. In recent years, viral misinformation on social media has been observed repeatedly, especially during elections and crisis situations \cite{Allcott.2017,Bakshy.2015,Pennycook.2020,Oh.2013}. In order to identify and eventually curb the spread of misinformation, experts fact-checkers on various third-party fact-checking organizations (\eg, \url{snopes.com}, \url{politifact.com}, \url{factcheck.org}) regularly investigate the veracity of social media rumors \cite{Wu.2019,Vosoughi.2018}. However, due to the limited amount of fact-checks that can be performed by these organizations, they are unable to accommodate the amount and speed of content creation on social media. Misinformation thus often continues to circulate and may only be detected when a tremendous amount of attention is paid to it \cite{Epstein.2020}. Furthermore, about 50\% of all Americans have concerns regarding the independence of the experts' assessment, \ie, distrust professional fact-checkers \cite{Poynter.2019}. Given these challenges, the real-world impact of fact-checks from third-party fact-checking organizations may be limited.


In order to address the drawbacks of the expert verification approach, recent research has proposed to employ non-expert fact-checkers in the crowd to verify social media content \cite{Micallef.2020,Bhuiyan.2020,Pennycook.2019,Epstein.2020,Allen.2020,Allen.2021,Godel.2021}. The rationale is that the ``wisdom of crowds'' (\ie, the aggregated assessments of non-expert fact-checkers) could result in an accuracy that is similar to that of experts \cite{Frey.2021}. Compared to the expert verification approach, harnessing the crowd for fact-checking would enable large numbers of fact-checks that could be carried out at higher frequency and lower cost \cite{Allen.2021,Pennycook.2019}. Furthermore, crowd-based fact-checking has the potential to remedy the problem of distrust in expert fact-checkers \cite{Allen.2021}. Recent experimental studies indeed yielded promising results -- suggesting that even relatively small crowds achieve an accuracy comparable to that of experts when fact-checking social media content \cite{Bhuiyan.2020,Epstein.2020,Pennycook.2019}.

While community-based fact-checking systems might be able to produce accurate fact-checks at scale, an understanding of how (mis-)information diffuses through social networks is still in its infancy. Prior works have analyzed the spread of rumors that have been fact-checked by third-party fact-checking organizations \cite{Friggeri.2014,Vosoughi.2018,Prollochs.2021b,Solovev.2022b}. For instance, several studies have compared characteristics of resharing cascades (\eg, how often a social media post is shared) across true vs. false rumors, finding that falsehood is more viral than the truth \cite{Vosoughi.2018,Prollochs.2021b,Solovev.2022b}. However, third-party fact-checking organizations tend to fact-check rumors on topics that are deemed to be of interest to a broad public and/or particularly concerning from the perspective of experts, while other misinformation remains unnoticed. In contrast, community fact-checked posts represent social media content that has been deemed worth fact-checking by actual social media users. Analyzing their diffusion would shed new light on the question of whether misinformation is more viral than the truth -- or rather a result of sample selection. However, we are not aware of any previous research analyzing the diffusion of crowd fact-checked posts on social media. Moreover, little is known about which social media posts are picked up in community-based fact-checking and how the spread varies across different types of misinformation (\eg, factual errors, missing context). Answering these questions is the goal of this study.


\textbf{Research questions: } In this work, we empirically analyze the diffusion of misleading vs. not misleading social media posts that have been fact-checked by the crowd. Specifically, we address the following research questions:
\begin{itemize}
	\item \textbf{(RQ1)} How do community fact-checked posts spread on social media? Are misleading posts more viral than not misleading posts?
	\item \textbf{(RQ2)} Are there differences in virality across different sub-types of community fact-checked misinformation (\eg, factual errors, missing context, manipulated media)?
	\item \textbf{(RQ3)} How do the fact-checking targets differ between community fact-checkers and expert fact-checkers?
	\item \textbf{(RQ4)} To what extent are (real-world) community-created fact-checks perceived as reliable?
\end{itemize}

\textbf{Data \& methodology:} We collect a comprehensive dataset consisting of community-created fact-checks from Twitter's Birdwatch platform. We then map the fact-checks to the fact-checked tweet using Twitter's historical API. This allows us to calculate the size of the resharing cascades (\ie, the number of retweets) in order to measure the virality of the fact-checked post. Subsequently, we implement an empirical regression model and link the fact-checking label to the number of retweets. We further control for the sentiment of the post and the social influence of its author (\eg, number of followers, account age, etc.). We then perform hypothesis testing to analyze whether posts categorized as being misleading are more viral than not misleading posts.

\textbf{Contributions:} This study is the first to analyze the spread of crowd fact-checked misinformation on social media. We show that crowd fact-checked misleading posts are \emph{less} viral than not misleading posts. Specifically, misleading posts are estimated to receive \SI{36.85}{\percent} fewer retweets than not misleading posts. Notably, this finding differs from earlier work \cite{Vosoughi.2018}, which has analyzed the diffusion of misinformation that has been fact-checked by third-party fact-checking organizations. We find that a partial explanation may lie in differences in the fact-checking targets: our findings suggest that community fact-checkers tend to fact-check posts from influential user accounts with many followers, while expert fact-checks tend to target rumors that are shared by less influential accounts. Our results further imply that there are significant differences in virality across different sub-types of misinformation (\eg, factual errors, missing context, manipulated media). 

As an additional contribution, we conduct a user study to assess the perceived reliability of (real-world) community-created fact-checks. Here, we observe that users agree with a large share of community-created fact-checks, whereas only a relatively small share is perceived as being purposely deceptive (e.g., due to motivated reasoning).

\section{Background}

\subsection{Misinformation on Social Media}

Over the last decade, the importance of social media (\eg, Twitter, Facebook) as an information platform for large parts of society has been subject to considerable growth \cite{Lazer.2018,Pew.2016}. On social media, any user can share information with his/her follower base \cite{Shore.2018}. Compared to traditional media, there is little control authority or oversight regarding the contents. For this reason, social media is highly vulnerable to the spread of misinformation. In fact, previous research suggests that social media platforms have become primary enablers of misinformation \cite[\eg,][]{Lazer.2018}. Online exposure to misinformation can affect how opinions are formed and causes detrimental societal effects \cite[\eg,][]{Allcott.2017,DelVicario.2016}. The latter has been repeatedly observed, especially during elections \cite[\eg,][]{Allcott.2017,Bakshy.2015} and crisis situations \cite[\eg,][]{Pennycook.2020b,Solovev.2022b,Oh.2013,Oh.2010,Starbird.2014}. 

A key feature of modern social media platforms is that users can also share others' content to increase its reach (\eg, ``retweeting'' on Twitter). This can result in misinformation cascades going ``viral.'' While previous research has mainly focused on characteristics and (negative) consequences of misinformation on social media, studies analyzing differences in the virality across misleading vs. not misleading posts are relatively scant. Existing works in this direction have analyzed the diffusion of posts that have been fact-checked by third-party fact-checking organizations \cite{Friggeri.2014,Vosoughi.2018,Solovev.2022b,Prollochs.2021b,Prollochs.2022b}. These studies found that misinformation diffuses significantly more virally than the truth. We are not aware of any previous study analyzing the spread of misleading vs. not misleading social media posts that have been fact-checked by the crowd. 

\subsection{Fact-Checking on Social Media}

Reliable fact-checking strategies are a crucial necessity to limit the spread of misinformation on social media. Currently, there are two predominant strategies. First, expert assessment in the form of human experts can check the veracity of content; \eg, via third-party fact-checking platforms (\eg, \url{snopes.com}, \url{politifact.com}, \url{factcheck.org}). Second, machine learning models can be trained to automatically classify misinformation \cite{Ma.2016,Qazvinian.2011}. For this purpose, content-based features (\eg, text, images, video), context-based features (\eg, time, location), or propagation patterns (\ie, how misinformation circulates among users) can be used. However, both methods suffer from several drawbacks. While experts classify misinformation fairly accurately, this strategy is difficult to scale due to the limited number of available humans experts \cite{Micallef.2022,Pennycook.2019}. Besides, a large proportion of social media users do not trust the independence of expert fact-checkers \cite{Poynter.2019}. In contrast, machine learning-based approaches are straightforward to scale, but typically show comparatively low accuracy \cite{Wu.2019}.

Given the trade-off between scalability and accuracy of existing approaches, recent works have proposed to outsource fact-checking of social media content to non-expert fact-checkers in the crowd \cite{Micallef.2020,Bhuiyan.2020,Pennycook.2019,Epstein.2020,Allen.2020,Allen.2021,Godel.2021}. The rationale is that the ``wisdom of crowds'' (\ie, the aggregated assessments of non-expert fact-checkers) could result in an accuracy that is comparable to that of experts \cite{Frey.2021,Woolley.2010}. The ability of crowds to ensure relatively trustworthy and high-quality accumulation of knowledge has been observed in various other online settings, such as on platforms like Wikipedia and Stack Overflow \cite[\eg,][]{Okoli.2014,Dissanayake.2019,Han.2021}. Applying a crowd-based approach to fact-check social media posts might have several benefits \cite{Pennycook.2019}. First, compared to expert assessments, significantly larger quantities of posts could be fact-checked. Second, trust issues with expert fact-checkers could, at least partially, be mitigated. Experimental studies suggest that, while the assessment of individuals might be noisy and ineffective \cite{Woolley.2010}, the crowd can be quite accurate in identifying misleading social media content. Here the assessment of even relatively small crowds has been found to be comparable to those of experts \cite{Bhuiyan.2020,Epstein.2020,Pennycook.2019}. Despite challenges with politically motivated reasoning \cite{Prollochs.2022a,Allen.2022}, recent research further shows that users, to a large extent, perceive community-created fact-checks for social media posts as being informative and helpful \cite{Prollochs.2022a}. 

\section{Data}

\subsection{Data Source: Community Fact-Checked Tweets from Birdwatch}\label{sec:birdwatch}

\begin{figure}[H]
	\centering
	\captionsetup{position=top}
	\subfloat[Misleading]{
		\begin{minipage}[b]{.4\textwidth}
			\centering
			{\fbox{\includegraphics[width=.7\linewidth]{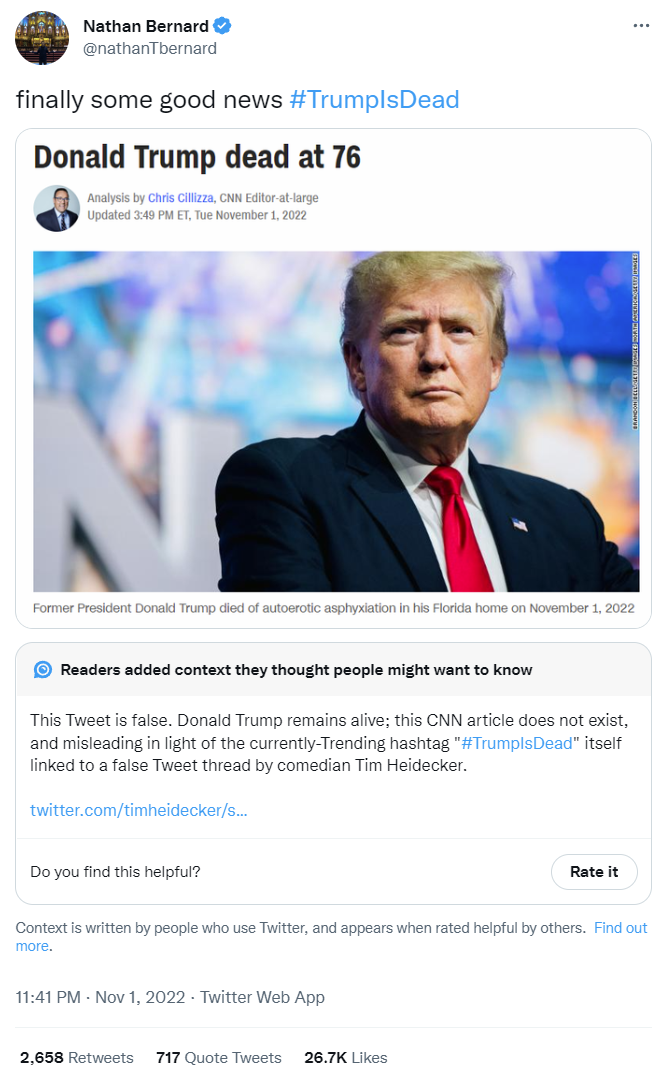}}\label{fig:screenshot_birdwatch_helpful_misleading}}  
	\end{minipage}}
	\hspace{0.35cm}
	\subfloat[Not misleading] {
		\begin{minipage}[b]{.4\textwidth}
			\centering
			{\fbox{\includegraphics[width=.7\linewidth]{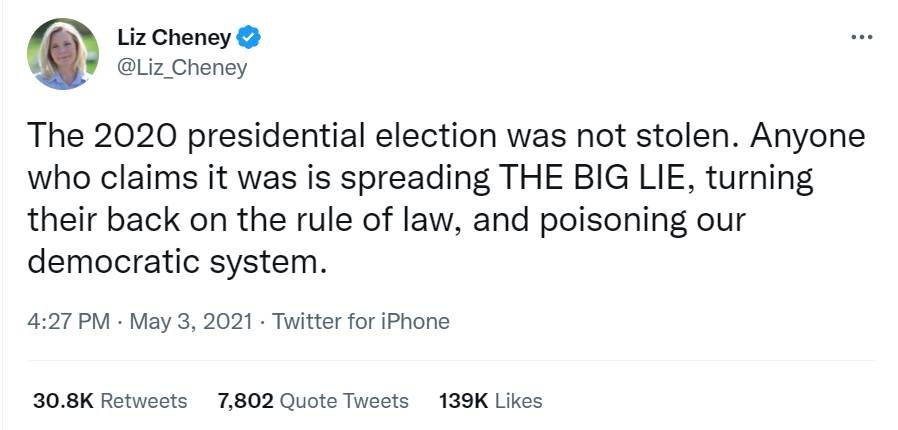}}\label{fig:screenshot_birdwatch_tweet_notmis}} 
	\end{minipage}}
	\caption{Examples of community fact-checked tweets. Only Birdwatch notes for misleading tweets are eligible to be directly shown on tweets. However, during our study period, community fact-checks from Birdwatch were practically not visible to the vast majority of social media users (\ie, only to pilot participants) and, thus, were unlikely to directly influence the diffusion of the fact-checked tweets. (a) Example of a tweet classified as misleading (\emph{Fact-Checking Label}) and the \emph{Text Explanation} of the corresponding Birdwatch note. The contributor selected ``Factual Error,'' ``Manipulated Media,'' and ``Satire'' as reasons for his/her classification (\emph{Misinformation Type}). (b) Example of a tweet classified as not misleading.
	}
	\label{fig:screenshot} 
	\vspace{-.5cm}
\end{figure}

We analyze the spread of social media posts that have been community fact-checked on Twitter's Birdwatch pilot \cite{Twitter.2021}. On January 23, 2021, Twitter launched Birdwatch as a new approach to address misinformation on their platform \cite{Twitter.2021}. The goal is to fact-check social media content by harnessing the ``wisdom of crowds.'' 
Birdwatch allows users to identify tweets they believe are misleading or not misleading and write notes that provide context to the tweet (so-called ``Birdwatch notes''). Users can fact-check \emph{any} tweet they come across on Twitter -- directly when browsing Twitter (see examples in Fig.~\ref{fig:screenshot}). Community fact-checking on Birdwatch comprises (1) checkbox questions that allow users to state whether a tweet might or might not be misleading (\emph{Fact-Checking Label}); (2) an open text field (max 280 characters) where users can explain their judgment (\emph{Text Explanation}), and (3) checkbox questions in which users can characterize the tweet and select reasons \emph{why} they perceive the tweet as being misleading (\emph{Misinformation Type}). For the latter, Birdwatch users can select one (or multiple) of the following answer options: (i) ``Factual Error,'' (ii) ``Missing Important Context,'' (iii) ``Unverified Claim as Fact,'' (iv) ``Outdated Information,'' (v) ``Manipulated Media,'' (vi) ``Satire,'' and (vii) ``Other.'' 

After a Birdwatch note is submitted, the fact-check is publicly available for other users to read. Birdwatch also features a rating system, which allows users to rate the helpfulness of the community-created fact-checks. These ratings are supposed to help identify which notes are most helpful and raise their visibility. Specifically, Birdwatch notes are shown directly on the fact-checked tweet if (i) the tweet is classified as misleading and (ii) it is rated by the community to be particularly helpful (see Fig.~\ref{fig:screenshot}).

Importantly, the data for this study originates from Birdwatch's pilot phase in the \US During this pilot phase, interested users were required to actively sign up to join Birdwatch. Any Twitter user could apply to become a Birdwatch contributor. Users that had signed up on Birdwatch could see Birdwatch notes directly when browsing Twitter next to the fact-checked tweet. Non-participating users could access Birdwatch notes via a separate Birdwatch website (\url{birdwatch.twitter.com}). In early 2022, Birdwatch had approximately 3250 contributors, compared to 41.5 million daily active Twitter users in the \US \cite{Statista.2022}.
Hence, during Birdwatch's pilot phase, community fact-checks from Birdwatch were practically not visible to the vast majority of social media users and, thus, were unlikely to directly influence the diffusion of the fact-checked tweets.\footnote{In early October 2022 (\ie, after our observation period), Twitter started to expand the Birdwatch program, allowing more Twitter users to view fact-checks directly on Twitter. Furthermore, Twitter rebranded Birdwatch to ``Community Notes.''} The Birdwatch pilot phase thus provides a unique opportunity to study the spread of community fact-checked posts with little confounding factors.

\subsection{Data Collection}

We downloaded \emph{all} Birdwatch notes between the introduction of the feature on January 23, 2021, and the end of February 2022 from the Birdwatch website, \ie, for an observation period of more than one year. The dataset contains a total number of \num{20218} Birdwatch notes (\ie, community-created fact-checks) from \num{3257} different contributors. We used the Twitter historical API to map the \textit{tweetID} referenced in each Birdwatch note to the source tweet. This approach allowed us to collect the following information about each source tweet and the account of its authors: (i) the number of retweets, (ii) the number of followers, (iii) the number of followees, (iv) the account age, and (v) whether the user has been verified by Twitter.

Notably, multiple Birdwatch users can write Birdwatch notes for the same tweet. Therefore, the data sometimes includes multiple fact-checks for the same post. The average number of Birdwatch notes per tweet is 1.33, with few tweets having many notes and most tweets having few. Only 18.79\% of the fact-checked tweets received more than one Birdwatch note. To avoid distortions due to multiple fact-checked tweets, we focus our analysis on the temporally first fact-check after the tweet has been posted. This filtering step resulted in a dataset consisting of \num{15256} unique fact-checks (for \num{15256} unique tweets). As part of our robustness checks, we also tested alternative approaches for handling multiple fact-checks (\eg, using Birdwatch's rating system, majority vote). Here we obtained qualitatively identical results.

\subsection{Variable Description}

Our dataset contains variables from two sources: (i) variables that are provided by the community-created fact-checks (\ie, the Birdwatch notes); and (ii) variables that represent information about the source tweet (\eg, the social influence of the author of the fact-checked tweet). 

\textbf{Fact-checks: } The Birdwatch notes provide us with the following variables: 
\begin{itemize}
	\item \textit{Misleading:} A binary indicator of whether a tweet has been reported as being misleading by the author of the Birdwatch note ($=1$; otherwise $=0$).
	\item \textit{Delay:} A numeric variable measuring the number of days elapsed between the posting date of the source tweet and the fact-check.
	\item \textit{Misinformation Type:} Seven dummy variables indicating reasons why a tweet has been reported as being misleading (``Factual Error,'' ``Missing Important Context,'' ``Unverified Claim as Fact,'' ``Outdated Information,'' ``Manipulated Media,'' ``Satire,'' and ``Other'').
\end{itemize}

\textbf{Source tweet: }
We used the Twitter historical API to map the \textit{tweetID} referenced in each Birdwatch note to the source tweet and collected the following information about each source tweet:
\begin{itemize}
	\item \textit{Retweet Count:} A numeric variable denoting the number of retweets a single tweet receives on Twitter. The retweet count is a common measure for the virality of a resharing cascade \cite[\eg,][]{Brady.2017,Solovev.2022b}.
	\item \textit{Followers:} The number of followers, \ie, the number of accounts that follow the author of the source tweet on Twitter.
	\item \textit{Followees:} The number of followees, \ie, the number of accounts whom the author of the source tweet follows on Twitter.
	\item \textit{Account Age:} The age of the author of the source tweet's account (in years).
	\item \textit{Verified:} A binary dummy indicating whether the account of the source tweet has been officially verified by Twitter ($=1$; otherwise $=0$).
	\item \textit{Sentiment:} We calculate a sentiment score measuring the positivity/negativity of the source tweet. Here we use a dictionary-based approach analogous to earlier research \cite[\eg,][]{Vosoughi.2018,Jakubik.2022,Baer.2022,Upworthy.2022,Rho.2020}. We first remove stopwords, punctuation, special characters (\eg, hashtags), and URLs in each source tweet. Subsequently, we employ the NRC lexicon \cite{Mohammad.2013}, which categorizes English words into positive and negative words. Following previous work \cite[\eg,][]{Rho.2020,Solovev.2022a}, the sentiment scores are then measured by calculating the difference between positive and negative words relative to the tweet length. For our sentiment analysis, we use the default implementation of the \texttt{sentimentr} package (with the built-in NRC lexicon) that also accounts for negations and valence shifters (see \cite{Rinker.2019} for details). 
\end{itemize}

\section{Empirical Analysis}

\subsection{Diffusion of Misleading vs. Not Misleading Posts (RQ1)}

We now empirically analyze the diffusion of misleading vs. not misleading posts that have been fact-checked on Twitter's Birdwatch platform. For this purpose, we first compare summary statistics. Note, however, that summary statistics should be interpreted with caution as the virality of social media posts strongly depends on the social influence of the author. To account for such confounding effects, we subsequently implement an empirical regression model with control variables that links the fact-checking label to the number of retweets. We then perform hypothesis testing to analyze whether posts categorized as being misleading are more viral than not misleading posts. 

\begin{figure}[H]
	\captionsetup{position=top}
	\captionsetup{belowskip=0pt}
	\centering
	\subfloat[Retweet Count]{\includegraphics[width=.28\linewidth]{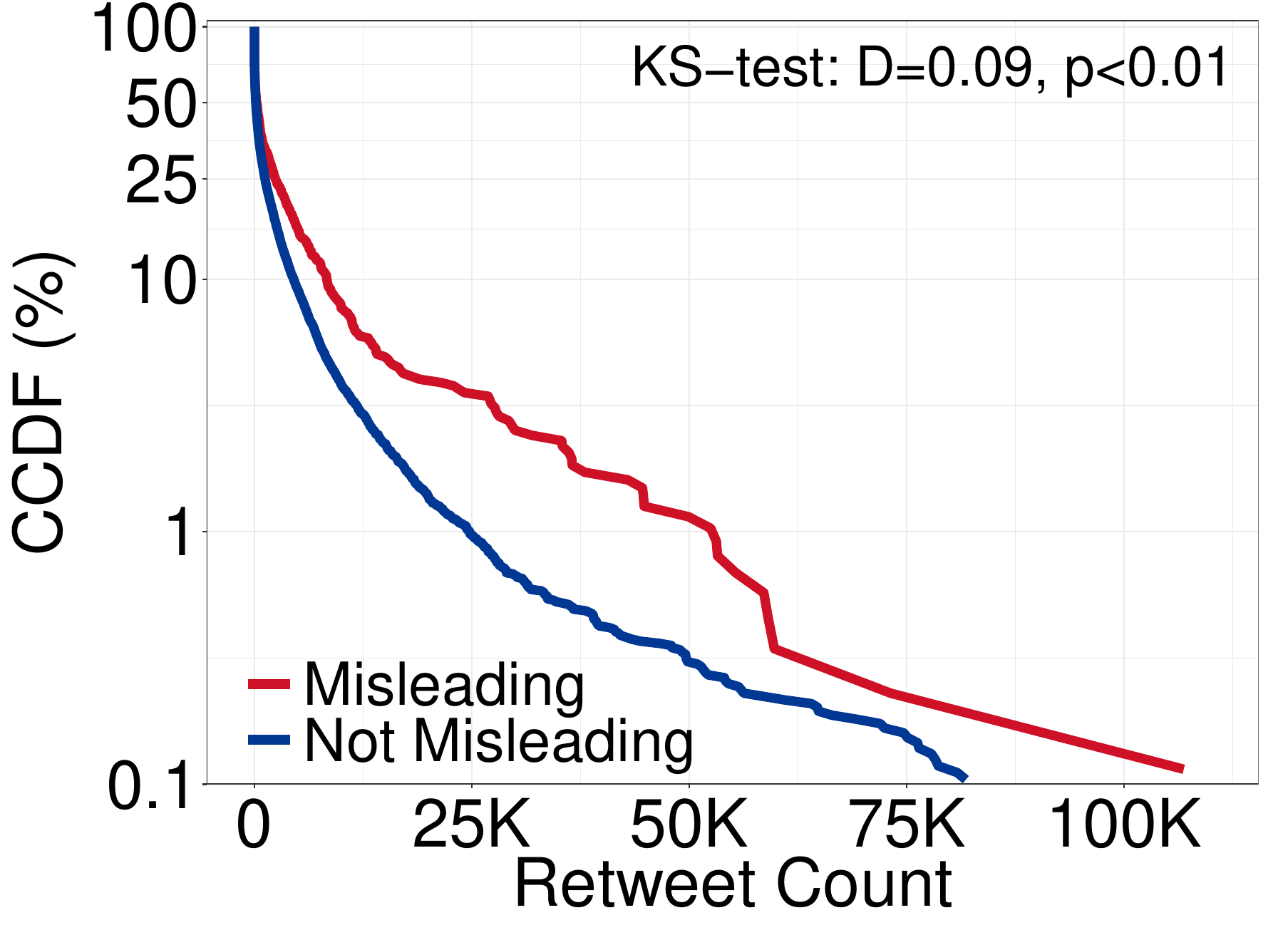}\label{fig:cascade_size}}
	\hspace{0.35cm}
	\subfloat[Sentiment]{\includegraphics[width=.28\linewidth]{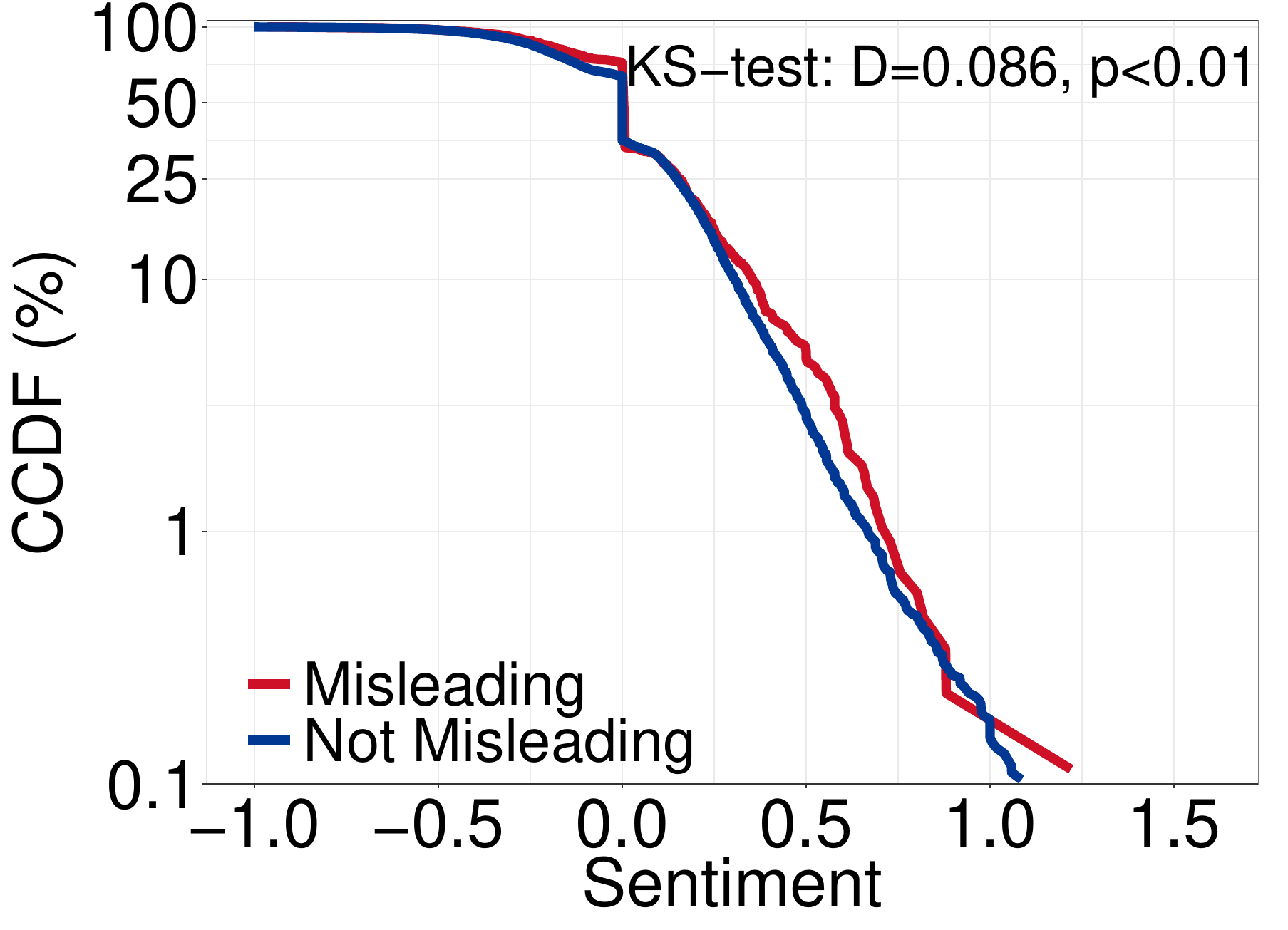}\label{fig:sentiment_source}}
	\hspace{0.35cm}
	\subfloat[Followers]{\includegraphics[width=.28\linewidth]{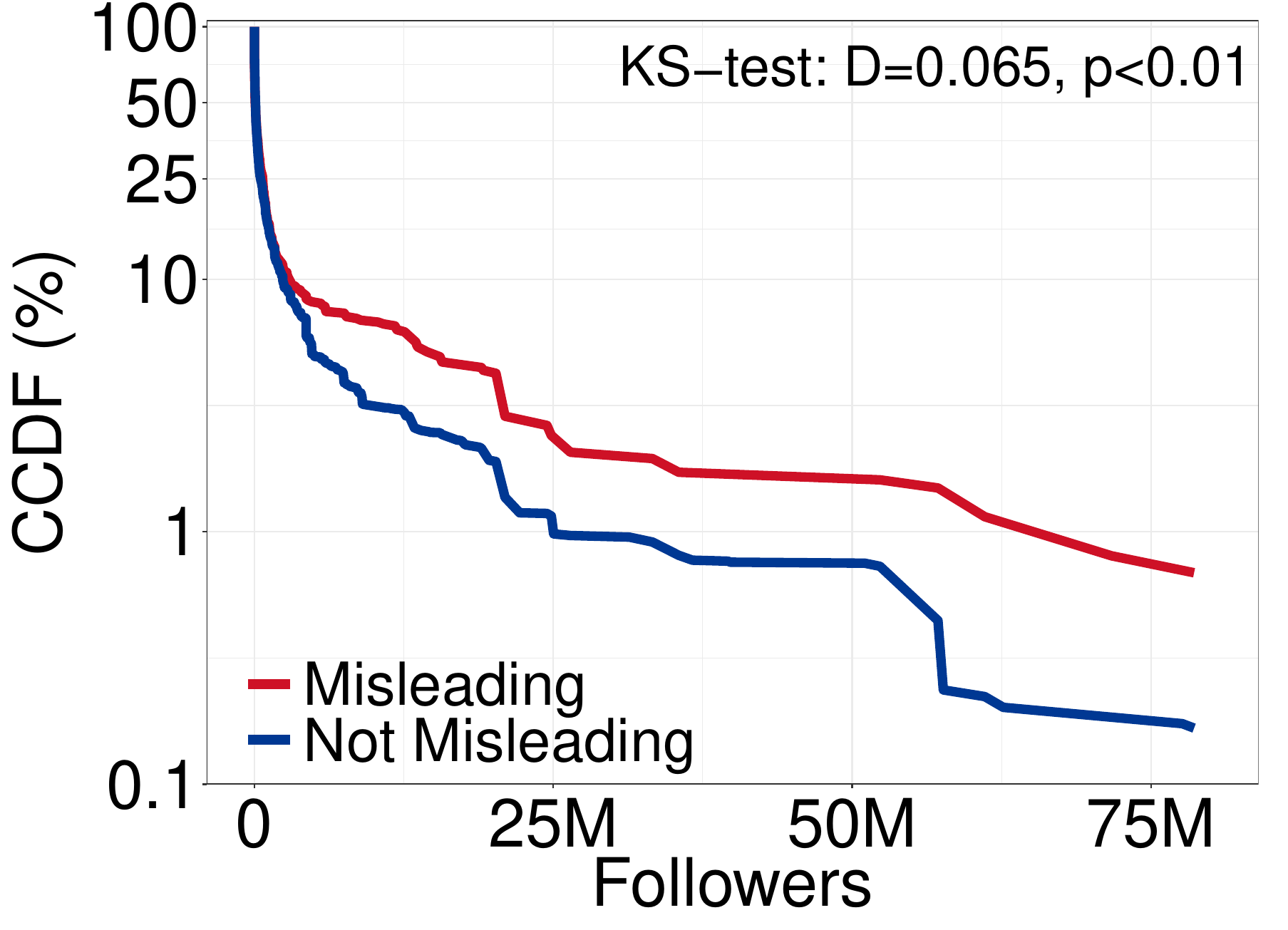}\label{fig:source_followers}} \\
	\subfloat[Followees]{\includegraphics[width=.28\linewidth]{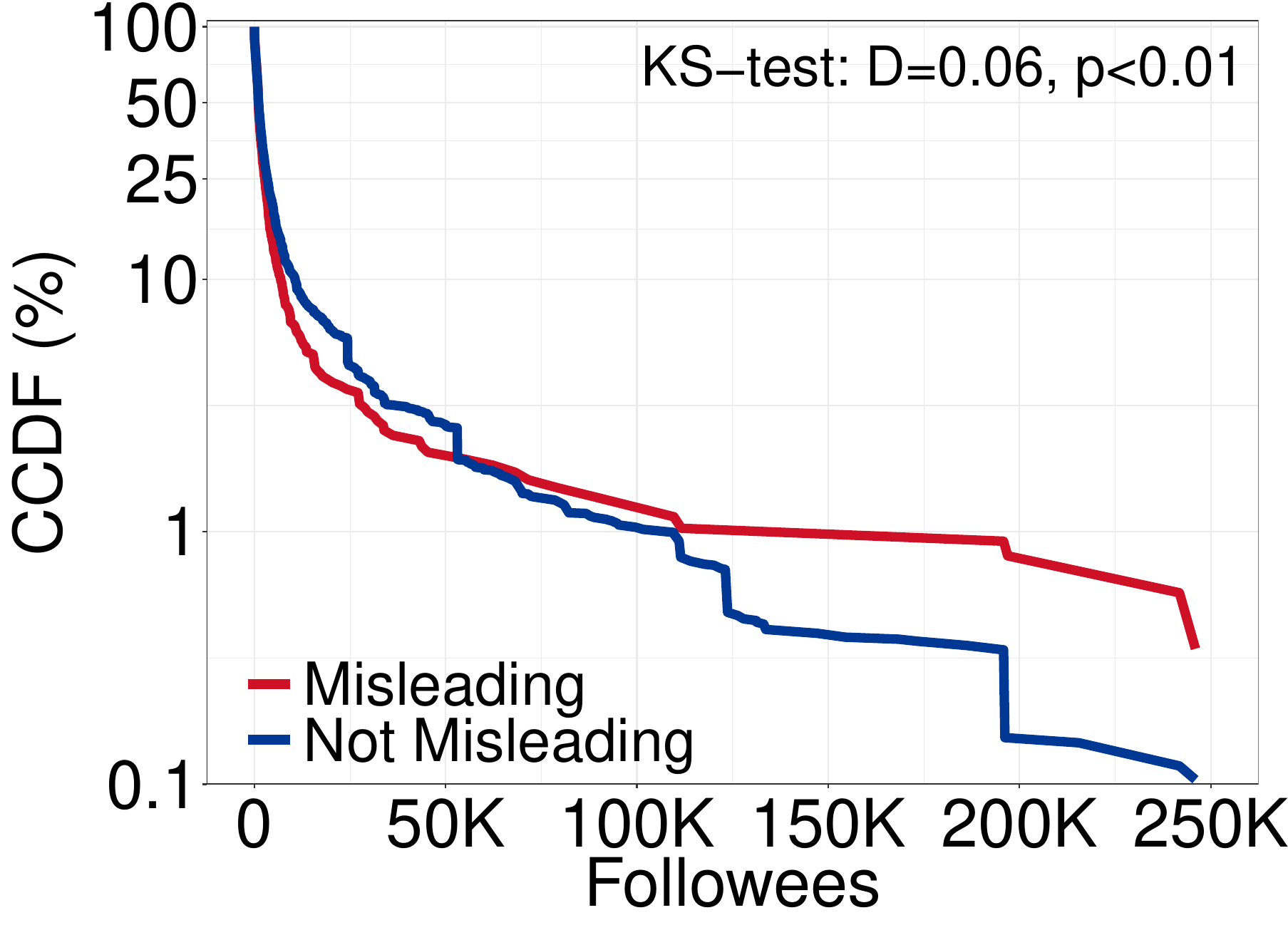}\label{fig:source_followees}}  
	\hspace{0.35cm}
	\subfloat[Account Age]{\includegraphics[width=.28\linewidth]{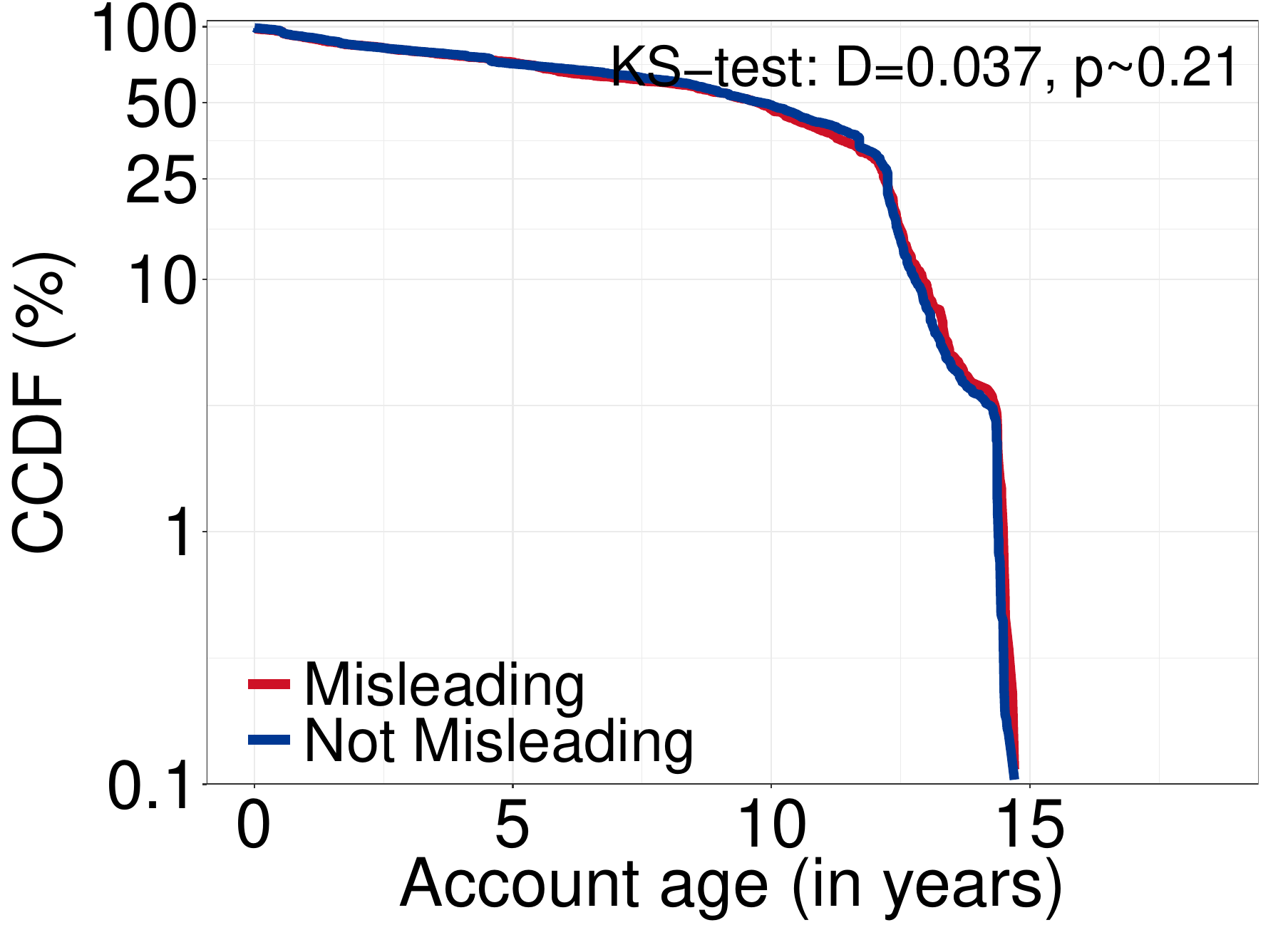}\label{fig:source_account_age}} 
	\hspace{0.35cm}
	\subfloat[Delay]{\includegraphics[width=.28\linewidth]{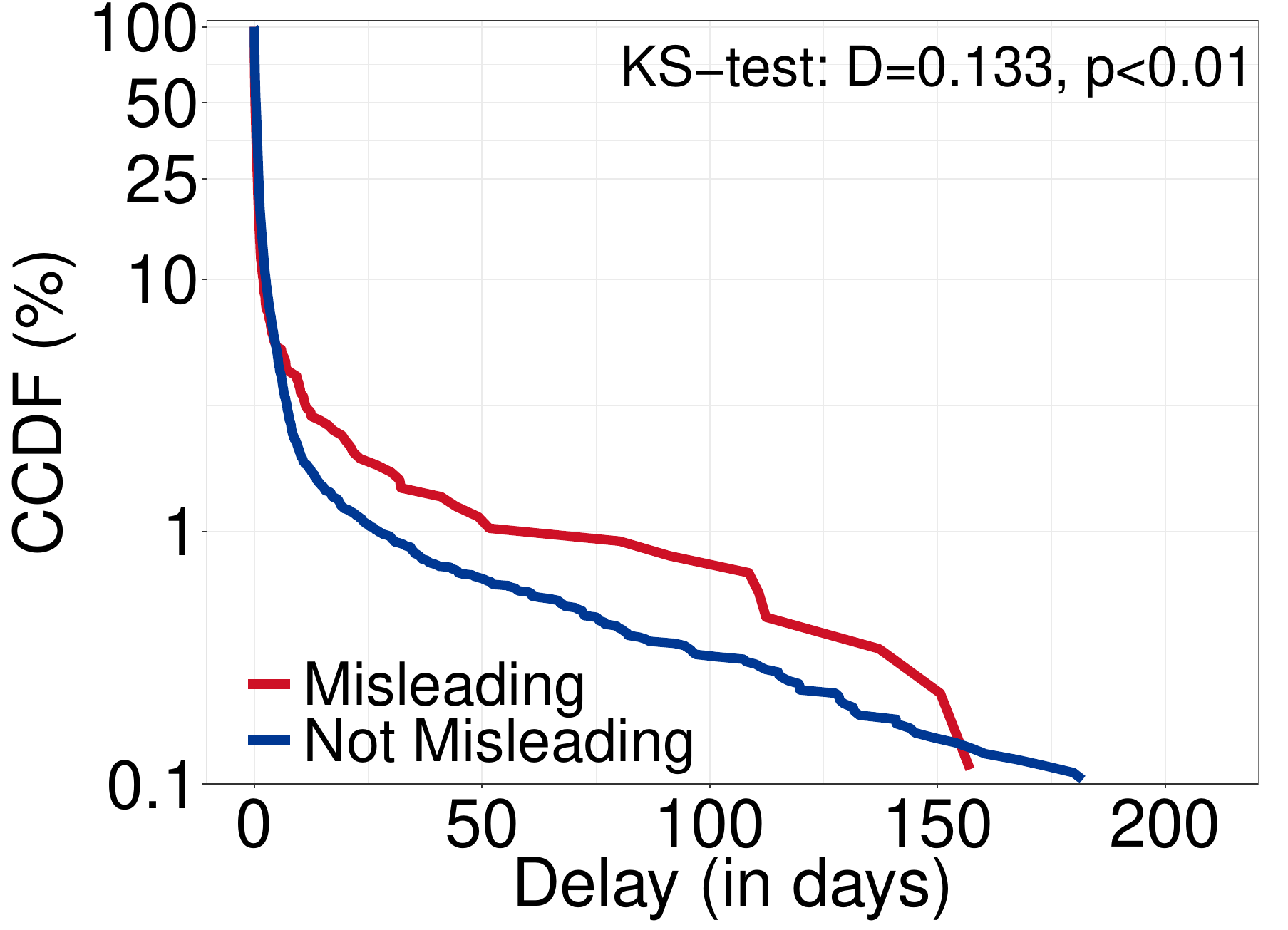}\label{fig:source_delay}}
	\caption{Complementary cumulative distribution functions (CCDFs) for (a) \emph{Retweet Count}, (b) \emph{Sentiment}, (c) \emph{Followers}, (d) \emph{Followees}, (e) \emph{Account Age}, and (f) \emph{Delay}.}
	\label{fig:summary_stats}
	\vspace{-.5cm}
\end{figure}

\textbf{Summary statistics:} Birdwatch users are vastly more likely to report misleading tweets than not misleading tweets. Out of \num{15256} community fact-checked tweets, \num{14384} (\SI{94.28}{\percent}) are classified as misleading and \num{872} (\SI{5.72}{\percent}) are classified as not misleading. In total, the fact-checked tweets in our dataset have been retweeted \num{29.45} million times. However, the retweet volume is higher for not misleading tweets than for misleading tweets. Specifically, the average retweets count amounts to \num{2478} for not misleading tweets and to \num{1478} for misleading tweets. A two-sided $t$-test confirms that the difference in means are statistically significant ($p<$ 0.01). Misleading vs. not misleading tweets also exhibit considerable heterogeneity with regards to sentiment and the social influence of the author. The sentiment tends to be significantly more positive in not misleading tweets (mean sentiment of \num{0.022}) than in misleading tweets (mean sentiment of \num{-0.004}). Misleading tweets are posted by users that have, on average, \SI{41.17}{\percent} fewer followers. 
Also here, two-sided $t$-tests confirm that the difference in means are statistically significant ($p<$ 0.01). We find only small differences in means for the variables \textit{Followees}, \textit{Account Age} and, \textit{Verified}, which are not statistically significant at common significance thresholds. Fig.~\ref{fig:summary_stats} further visualizes the complementary cumulative distribution functions (CCDFs). Kolmogorov-Smirnov (KS) tests show that, with the exception of \textit{Account Age}, the differences in the distributions between misleading and not misleading tweets are statistically significant ($p<0.01$).  

\textbf{Regression model:} We implement explanatory regression analysis to better understand the diffusion of misleading vs. not misleading crowd fact-checked posts. In contrast to summary statistics, this allows us to estimate effect sizes after controlling for confounding effects. The dependent variable in our regression analysis is given by $\mathit{Retweet Count_i}$, that is, the number of retweets for a fact-checked tweet \textit{i}. The retweet count is a non-negative count variable, and its variance is larger than the mean. To adjust for overdispersion, we draw upon a negative binomial regression to model the retweets count \cite{Solovev.2022b,Prollochs.2021b}. The key explanatory variable is $\mathit{Misleading_i}$, \ie, whether the tweet has been classified as misleading by Birdwatch users (\ie, $=1$ if true, otherwise $=0$). Additionally, we include the elapsed time between the publication of the tweet and the fact-check ($\mathit{Delay_i}$). Furthermore, we must control for the social influence of the source tweet and its author. Therefore, we adjust for variables known to affect the retweet rate \cite{Stieglitz.2013,Brady.2017,Vosoughi.2018,Solovev.2022b,Prollochs.2021a}, which includes the number of followers ($\mathit{Followers}_i$) and followees ($\mathit{Followees}_i$), the account age ($\mathit{AccountAge}_i$), and whether the account was verified by Twitter ($\mathit{Verified}_i$). In addition, we control for the sentiment of the source tweet ($\mathit{Sentiment}_i$). The resulting model is

{\footnotesize
	\begin{align}
		\log(&{\mathup{E}(RetweetCount_i \,\mid\, ^*)})  = \,\beta_{0}  +  \beta_{1} \, \mathit{Misleading}_i \label{eq:neg_bin} \\
		          &  + \beta_{2} \, \mathit{Delay}_i + \beta_{3} \,  \mathit{Sentiment}_i  + \beta_{4} \, \mathit{Followers}_i \nonumber \\
		          &  + \beta_{5} \, \mathit{Followees}_i + \beta_{6} \, \mathit{Account Age}_i + \beta_{7} \, \mathit{Verified}_i + u_{i}, \nonumber 
	\end{align} 
}%
\normalsize

with intercept $\beta_0$ and month-year fixed effects $u_{i}$ to adjust for differences in the start date and age of the resharing cascades. For the sake of interpretability, we $z$-standardize all continuous variables. This allows us to compare the effects of regression coefficients on the dependent variable measured in standard deviations. Note that since we apply a negative binomial regression, the interpretation of the effect sizes requires an exponential
transformation of the coefficients.

\begin{figure}[H]
		\captionsetup{position=top}
		\captionsetup{belowskip=0pt}
		\captionsetup[subfloat]{textfont={sf,normalsize}, skip=2pt, singlelinecheck=false, labelformat=simple,labelfont=bf,justification=centering}
		\centering
		\includegraphics[width=.5\linewidth]{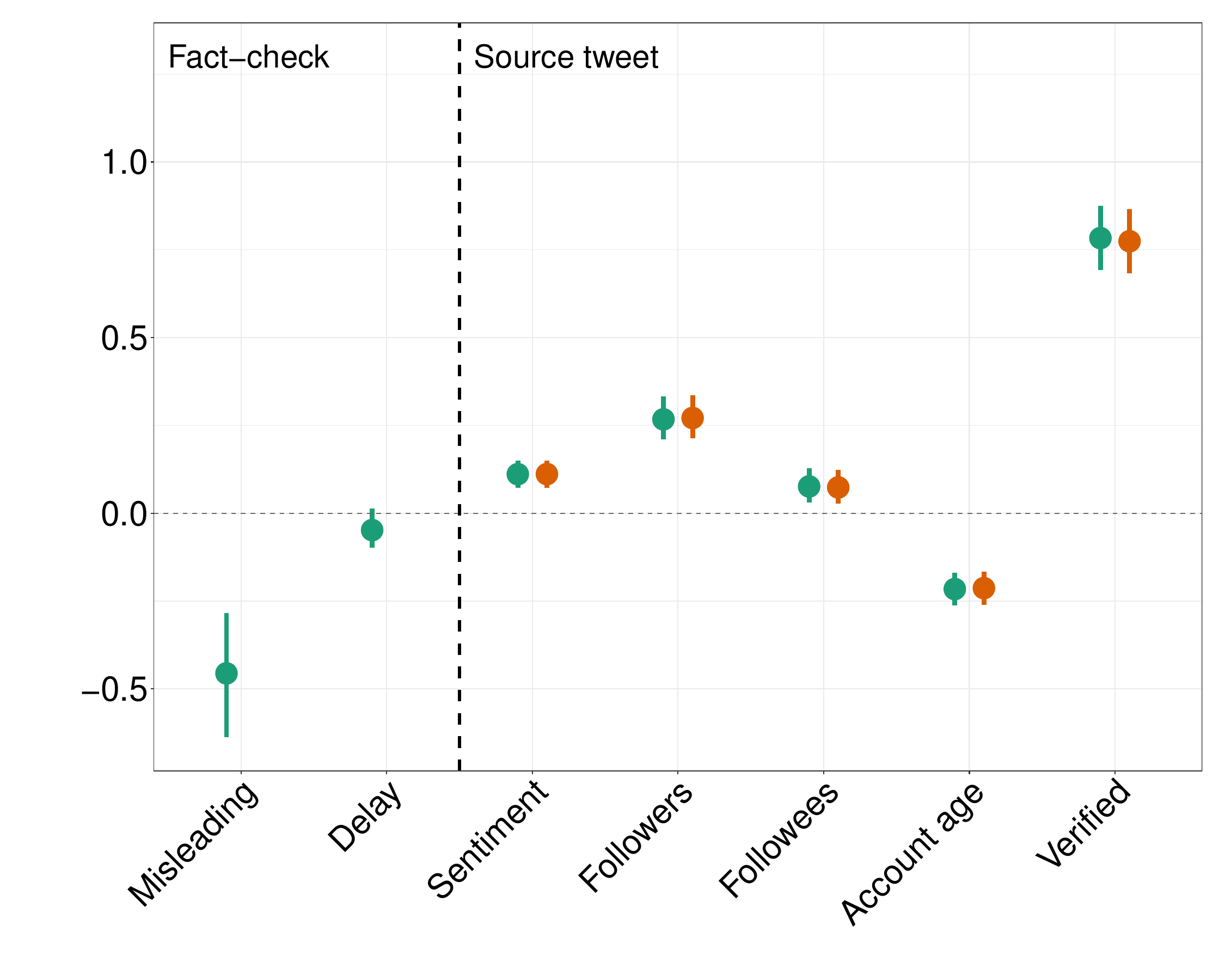}
		\caption{Coefficient estimates for negative binomial regression with the retweet count as dependent variable. Model (a) includes all variables given by the source tweet (orange). Model (b) additionally includes variables concerning the fact-check (green). The vertical bars represent \SI{99}{\percent} confidence intervals. Month-year fixed effects are included.} 
		\label{fig:main_model}
		\vspace{-.5cm}
\end{figure}

\textbf{Coefficient estimates:} The coefficient estimates for the regression model are reported in Fig.~\ref{fig:main_model}. We find that misleading tweets are significantly less viral than not misleading tweets. Specifically, the coefficient for $\mathit{Misleading}$ is \num{-0.456} ($p<0.01)$, which implies that misleading tweets are expected to receive $e^{-0.459}-1 \approx$ \SI{36.62}{\percent} fewer retweets. Furthermore, we observe that the coefficient estimate for $\mathit{Delay}$ is small in magnitude and not statistically significant at common significance threshold. This implies that differences in the fact-checking speed are not significantly associated with differences in virality of crowd fact-checked posts. 

Concordant with the literature \cite{Stieglitz.2013,Solovev.2022b,Vosoughi.2018}, we observe statistically significant estimates for the variables characterizing the social influence of the author of the source tweet. The number of followers has a large positive effect on the number of retweets (coef: $0.267$; $p<0.01$), while the number of followees has a smaller positive effect (coef: $0.076$; $p<0.01$). A higher account age decreases the expected number of retweets (coef: $-0.216$; $p<0.01$), while posts from verified accounts are expected to receive more retweets (coef: $0.783$; $p<0.01$). Similar to earlier work \cite{Prollochs.2021b}, we also find that more positive sentiment is associated with more retweets (coef: $0.111$; $p<0.01$). 

\subsection{Diffusion of Different Types of Misinformation (RQ2)}

\begin{figure}[H]
		\captionsetup{position=top}
		\captionsetup{belowskip=0pt}
		\captionsetup[subfloat]{textfont={sf,normalsize}, skip=2pt, singlelinecheck=false, labelformat=simple,labelfont=bf,justification=centering}
		\centering
		\includegraphics[width=.5\linewidth]{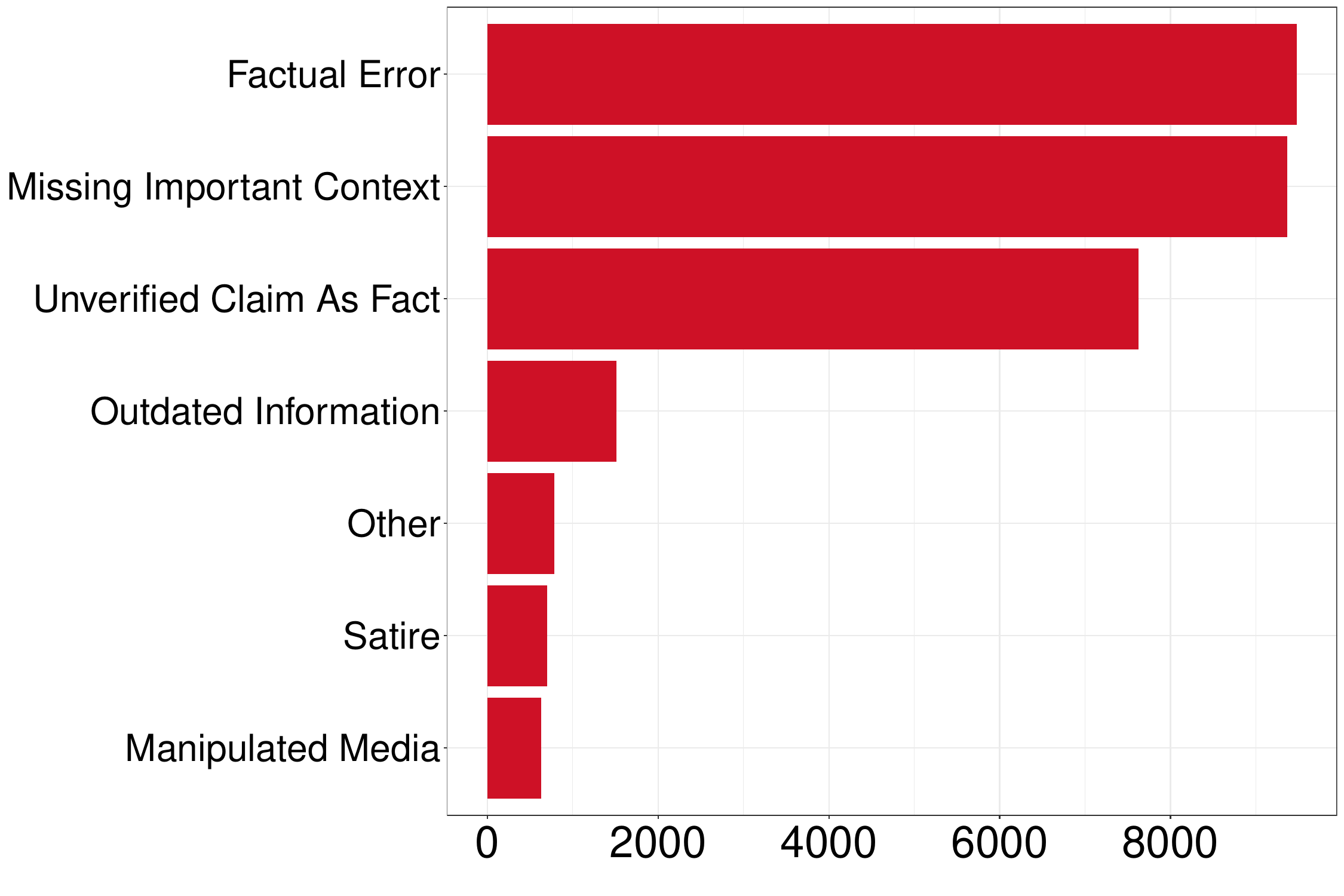}
		\caption{Barplot showing the number of tweets per checkbox answer option in response to the question ``Why do you believe this tweet may be misleading?''}
		\label{fig:frequency_subtypes_misinformation}
		\vspace{-.5cm}
\end{figure}

If fact-checkers on Birdwatch have classified a tweet as being misleading, they additionally need to answer checkbox questions on the reasons \emph{why} they perceive it as such. As aforementioned, Birdwatch users can select one (or multiple) of the following answer options: (i) ``Factual Error,'' (ii) ``Missing Important Context,'' (iii) ``Unverified Claim as Fact,''  (iv) ``Outdated Information,'' (v) ``Manipulated Media,'' (vi) ``Satire,'' and (i) ``Other.'' Fig.~\ref{fig:frequency_subtypes_misinformation} shows that the vast majority of tweets have been categorized as misleading because of factual errors (\SI{62.13}{\percent}), missing context (\SI{61.38}{\percent}), or because they treat unverified claims as fact (\SI{49.99}{\percent}). The other categories are relatively rare. 

\begin{figure}[H]
		\captionsetup{position=top}
				\captionsetup{belowskip=0pt}
		\captionsetup[subfloat]{textfont={sf,normalsize}, skip=2pt, singlelinecheck=false, labelformat=simple,labelfont=bf,justification=centering}
		\centering
		\includegraphics[width=.5\linewidth]{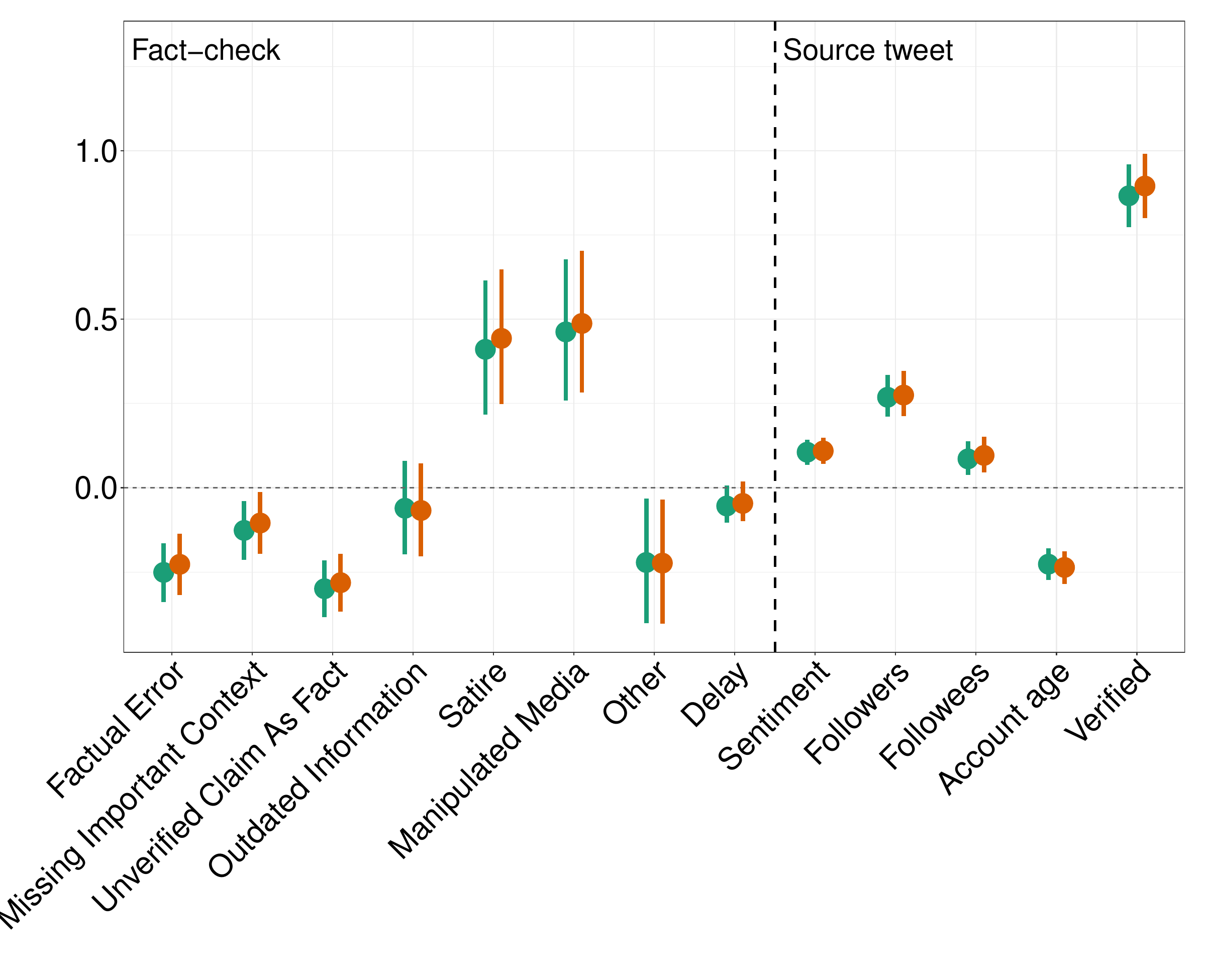}
		\caption{Coefficient estimates for negative binomial regression with the retweet count as dependent variable. Here, dummy variables referring to different sub-types of misinformation are included. Model (a) includes all posts (green), whereas Model (b) only includes the subset of posts classified as misleading (orange). The reference type in Model (a) are tweets classified as ``not misleading,'' whereas the reference type in Model (b) are misleading tweets that have not been assigned to a subtype.
			The vertical bars represent \SI{99}{\percent} confidence intervals. Month-year fixed effects are included.}
		\label{fig:model_subtypes}
		\vspace{-.5cm}
\end{figure}

We repeat our regression analysis with dummy variables referring to the different types of misleading posts as provided by Birdwatch contributors. This allows us to examine differences in the virality across different types of misinformation. The coefficient estimates in Fig.~\ref{fig:model_subtypes} show that misleading tweets are less viral than not misleading tweets if they belong to the misinformation sub-types ``Factual Error'' (coef: $-0.251$; $p<0.01$), ``Missing Important Context'' (coef: $-0.127$; $p<0.01$), ``Unverified Claim as Fact'' (coef: $-0.300$; $p<0.01$) and, ``Other'' (coef: $-0.221$; $p<0.01$). In contrast, tweets belonging to the misinformation sub-types ``Manipulated Media'' (coef: $0.461$; $p<0.01$), and ``Satire'' (coef: $0.411$; $p<0.01$) receive more retweets. These results suggest that there are significant differences in virality across different sub-types of misinformation. The coefficient estimates for the other variables do not differ qualitatively from the previously performed regressions. 

\newpage
\subsection{Comparison to Expert-Based Fact-Checking (RQ3)}

In contrast to the work by \citeauthor{Vosoughi.2018}~(\citeyear{Vosoughi.2018}), which found that expert fact-checked falsehood on Twitter is \emph{more} viral than the truth, our analysis suggests that crowd fact-checked tweets perceived as misleading are \emph{less} viral than those perceived as not misleading. A possible explanation for this finding lies in the sample selection, \ie, third-party fact-checking organizations vs. Birdwatch contributors might fact-check social media posts published by different account types.

\begin{figure}[H]
	\captionsetup{position=top}
			\captionsetup{belowskip=0pt}
	\centering
	\subfloat[Followers]{\includegraphics[width=.4\linewidth]{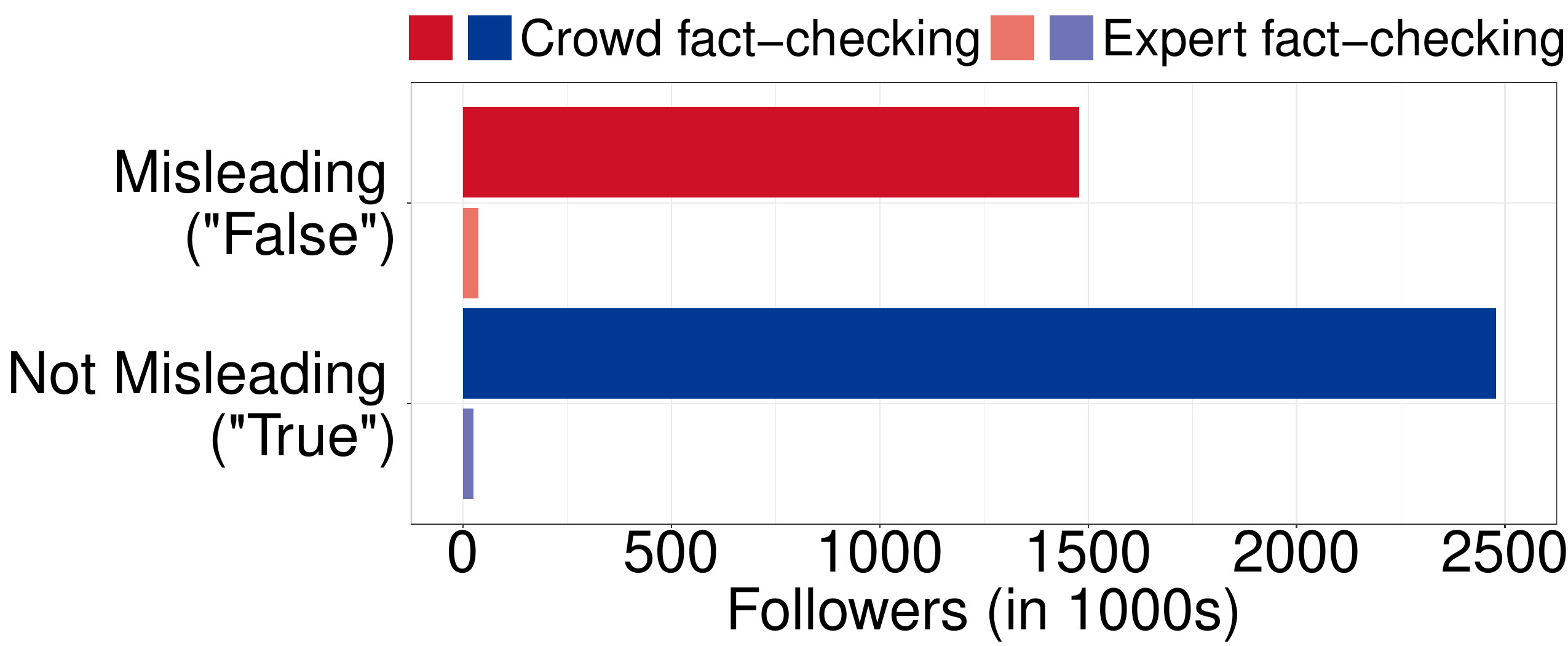}\label{fig:comparison_followers}} 
	\hspace{0.15cm}
	\subfloat[Followees]{\includegraphics[width=.4\linewidth]{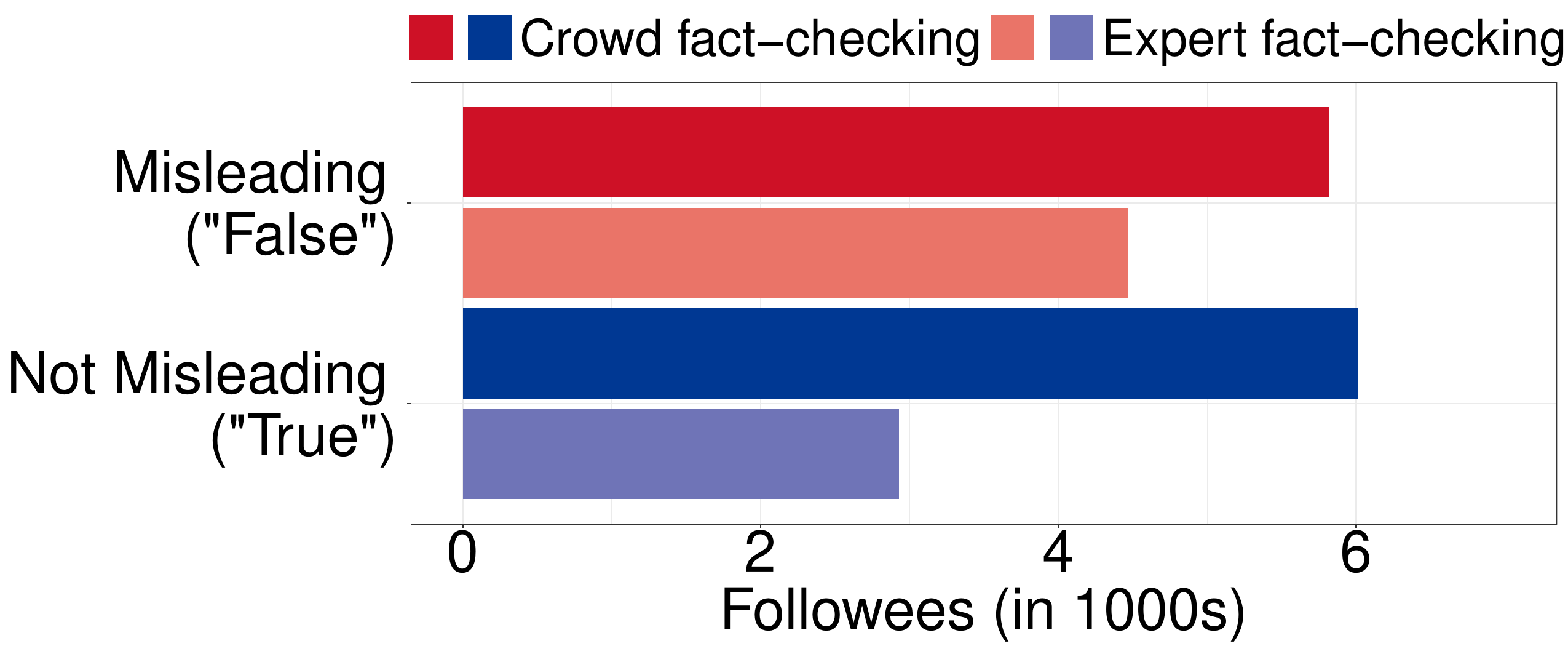}\label{fig:comparison_followees}} \\
	\subfloat[Account Age]{\includegraphics[width=.4\linewidth]{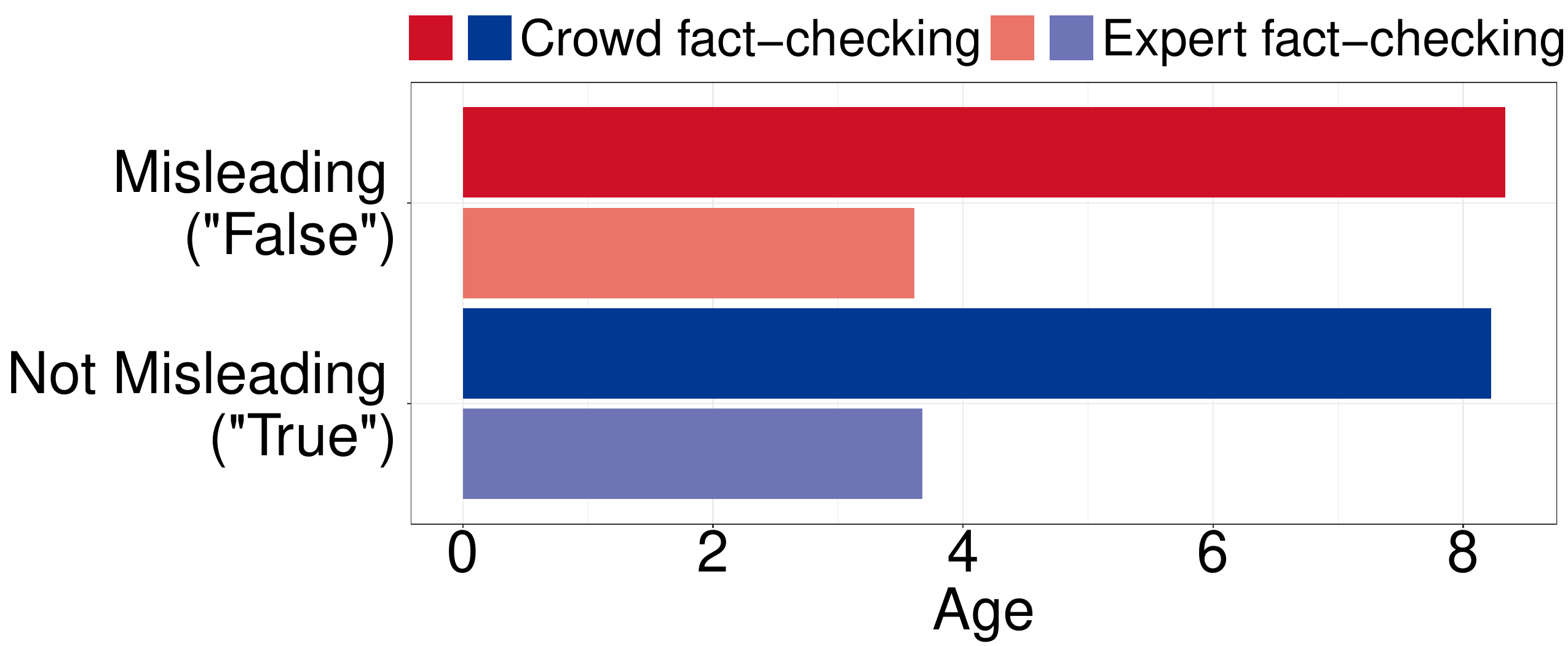}\label{fig:comparison_age}}
	\hspace{0.15cm}
	\subfloat[Verified]{\includegraphics[width=.4\linewidth]{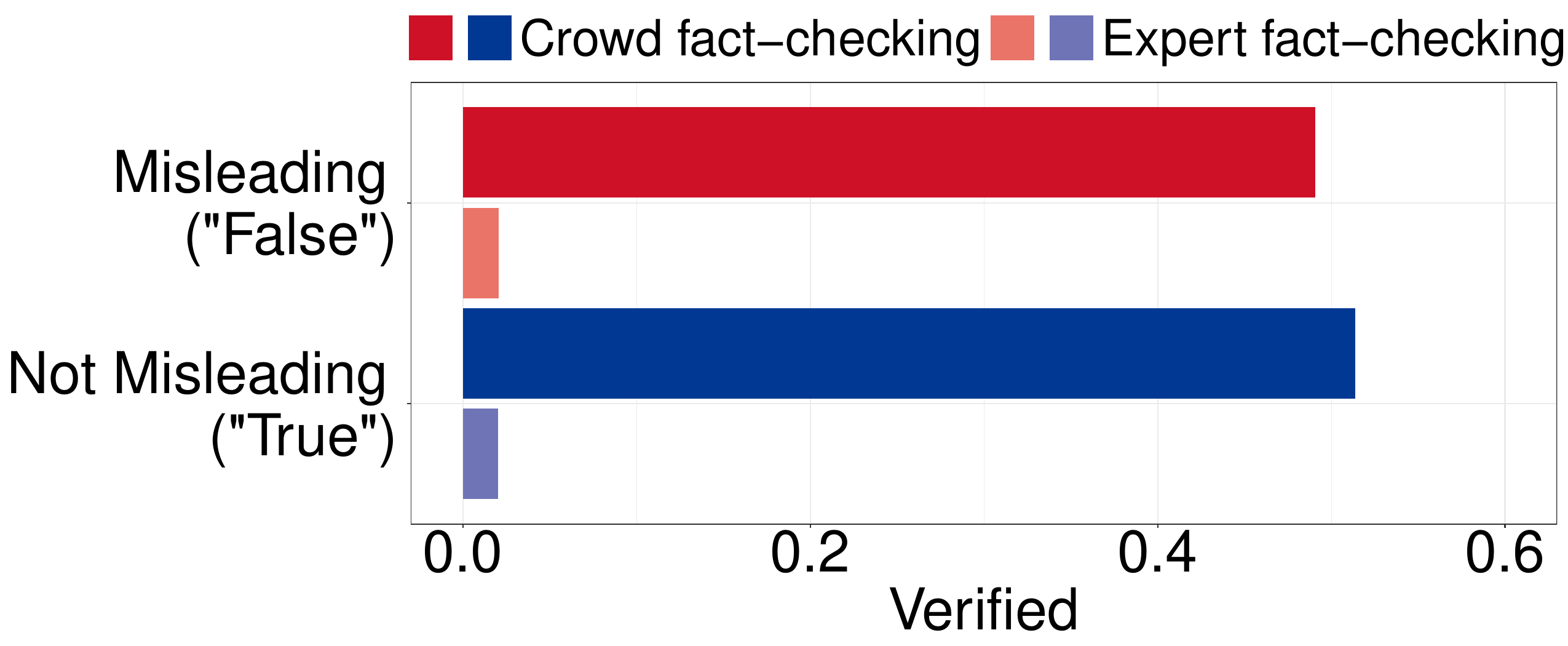}\label{fig:comparison_verified}} \\
	\caption{Comparison of characteristics (mean values) of authors of crowd fact-checked and expert fact-checked tweets for (a) the number of followers, (b) the number of followees, (c) the account age and, (d) the verified status. We compare the authors of misleading and not misleading crowd fact-checked posts on Birdwatch to those of true and false rumors in the dataset of expert fact-checked posts from \citeauthor{Vosoughi.2018}~(\citeyear{Vosoughi.2018}).
	}
	\label{fig:comparison_vosoughi}
	\vspace{-.5cm}
\end{figure}

To shed light on this question, Fig.~\ref{fig:comparison_vosoughi} compares the mean values of different user characteristics of the authors of misleading and not misleading crowd fact-checked posts to those of authors true and false rumors in the dataset of expert fact-checked posts from \citeauthor{Vosoughi.2018}~(\citeyear{Vosoughi.2018}). Compared to expert fact-checked tweets, we find that user accounts of authors of crowd fact-checked posts have, on average, $\approx$~40 times more followers, \SI{41.65}{\percent} more followees, and approximately twice the account age. Moreover, while \SI{49.21}{\percent} percent of the accounts of authors of crowd fact-checked posts are verified by Twitter, this is only the case for \SI{2.00}{\percent} of the authors of expert fact-checked posts. Two-sided $t$-tests confirm that each difference in means is statistically significant ($p<$ 0.01). These findings suggest that social media users contributing to crowd-based fact-checking tend to fact-check posts from larger accounts with greater social influence, while expert fact-checks tend to target rumors that are shared by smaller accounts.

We further observe that, for expert fact-checked posts, falsehood tends to originate from accounts with relatively more followers, while we observe the opposite pattern for crowd fact-checked posts (see Fig.~\ref{fig:comparison_vosoughi}). Specifically, we find that authors of crowd fact-checked posts perceived as misleading have \SI{67.71}{\percent} more followers than accounts of posts perceived as not misleading ($p<$ 0.01). In contrast, authors of falsehood in expert fact-checked posts have \SI{34.04}{\percent} less followers than authors of the true tweets ($p<$ 0.01). This suggests that fact-checks from Birdwatch contributors are more likely to endorse/emphasize the accuracy of not misleading tweets authored by influential users with a wide reach. Opposite to this, expert fact-checked tweets authored by influential accounts are more likely to convey false information. 
Since author characteristics are inherently linked to the virality of posts (\eg, users with a wider reach can generate more retweets), the observed differences in fact-checking targets provide a (partial) explanation for the overall higher virality of not misleading posts in the case of Birdwatch. 

\subsection{Perceived Reliability of Community-Created Fact-Checks (RQ4)}

In order to assess the perceived reliability of the community-based fact-checks from Birdwatch, we conducted a user study on the online survey platform Prolific (\url{www.prolific.com}). We recruited \textit{n = 7} participants, four women and three men, who were on average 35 years old. All participants were based in the \US, and English native speakers. All but one participant indicated that they are familiar with Twitter and regularly share content on social media.
Participants were presented with a randomized sample of \num{300} tweets (\num{150} not misleading and \num{150} misleading) and the corresponding fact-checks from Birdwatch (fact-checking label and text explanation). Note that we purposely presented the participants with both the source tweet and the fact-check (instead of only the source tweet). In the absence of a ground truth (which might require expert assessment), we were interested in the \emph{perceived} reliability of the fact-checks rather than testing how much one crowd agrees with another. As such, for each tweet, participants were asked for their assessment on (i) the extent to which they agree with the fact-checking label, and (ii) whether they perceive the fact-check as purposely deceptive (\eg, because of motivated reasoning, manipulation attempts, etc.). The participants answered both questions on a 5-point Likert scale, ranging from 1 (``strongly disagree'') to 5 (``strongly agree''). 

\begin{figure}[H]
	\captionsetup{position=top}
			\captionsetup{belowskip=0pt}
	\centering
	\subfloat[Agreement with Fact-Checks]{\includegraphics[width=.45\linewidth]{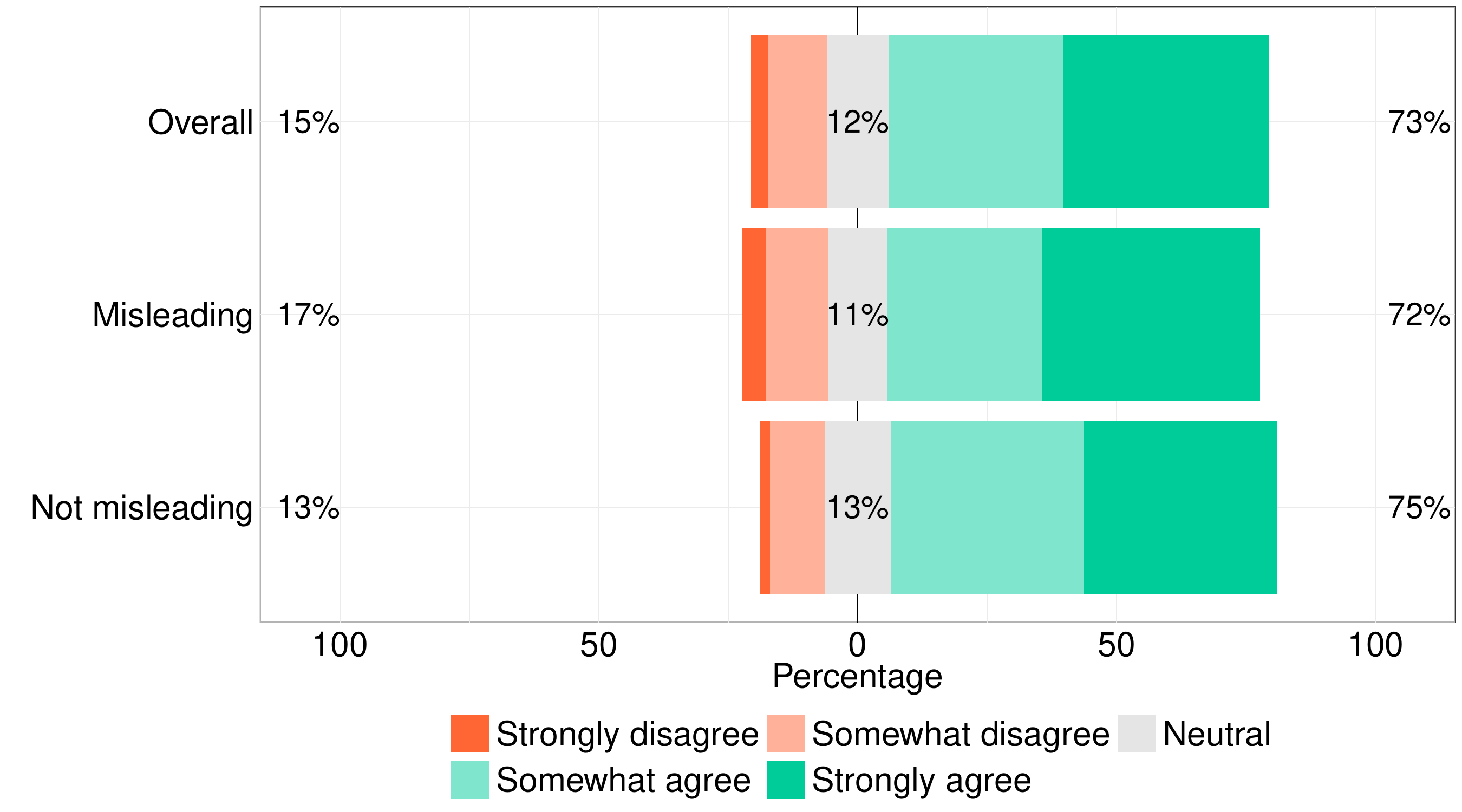}\label{fig:shares_agreement}} 
	\hspace{0.15cm}
	\subfloat[Purposely Deceptive Fact-Checks]{\includegraphics[width=.45\linewidth]{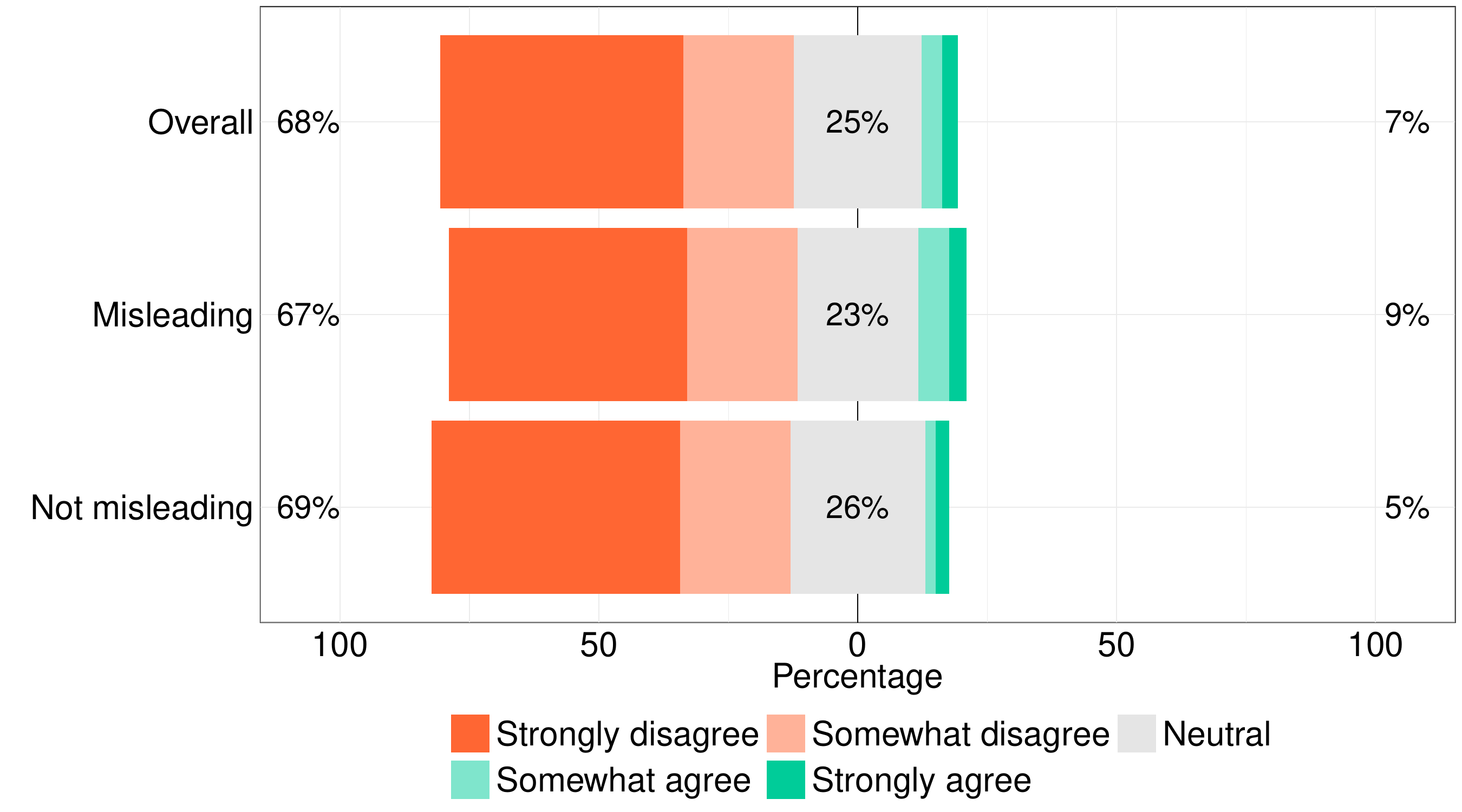}\label{fig:shares_purpose}} \\
	\caption{User study evaluating the perceived reliability of community-based fact-checks from Birdwatch. $n = 7$ participants were recruited via Prolific. Here we report the median responses to the questions (a) ``Do you agree with the fact-checking label?'' and (b) ``Do you feel that the fact-check is purposely deceptive?''}
	\label{fig:shares_userstudy}
	\vspace{-.5cm}
\end{figure}

Fig.~\ref{fig:shares_userstudy} visualizes the distribution of the median votes for the individual tweets across all response options. We first evaluate the extent to which the participants agree with the fact-checks from Birdwatch. We find that the participants at least somewhat agree with \SI{73.33}{\percent} of the community-created fact-checks performed by Birdwatch users. Interestingly, the agreement is lower for tweets categorized as misleading (\SI{72.00}{\percent}) than for tweets classified as not misleading (\SI{74.67}{\percent}). 

We find a consistent pattern for the second question item: the median ratings of the seven participants suggest that only a relatively small share of \SI{7.00}{\percent} of fact-checks are perceived as purposely deceptive. Notable, we again observe considerable differences across fact-checks across fact-checks reporting misleading vs. not misleading tweets. Specifically, fact-checks reporting misleading tweets are more likely to be perceived as being purposely deceptive (\SI{9.33}{\percent}) than fact-checks reporting not misleading tweets (\SI{4.67}{\percent}).

The participants showed statistically significant inter-rater agreements. Kendall's $W$ was \num{0.43} ($p<0.01$) for the first question item (agreement with the fact-checking label); and \num{0.32} ($p<0.01$) for the second question item (purposely deceptive fact-checks). 

In sum, the results of our user study suggest that the vast majority of community-created fact-checks are perceived as being reliable. This supports the results of previous experimental works, which suggest that the risk of users purposely trying to ``game the system'' is tolerable \cite[\eg,][]{Allen.2021}. Even though inaccurate fact-checks and misuse of the platform cannot be prevented completely, community-based fact-checking should be seen as one tool (as part of a larger toolset) that may help to combat the spread of misinformation on social media \cite{Epstein.2020,Godel.2021}.

\subsection{Robustness Checks} 
We conducted an extensive set of checks that yielded consistent findings: (1)~we controlled for outliers in the dependent variables; (2)~we ran separate regressions for misleading vs. not misleading posts; (3)~we calculated variance inflation factors for all independent variables and found that all remain below the critical threshold of four; (4) we repeated our analysis with user-specific random effects; (5)~we incorporated quadratic effects; (6)~we included interaction terms between user-specific variables and the fact-checking label; (7)~we evaluated alternative approaches to handle multiple fact-checks for the same tweet (\eg, majority vote, no filtering of multiple fact-checks). In all of these checks, our findings are supported. Detailed results are reported in the supplementary materials.

\section{Discussion}

\textbf{Summary of findings:} This study is the first to examine the diffusion of misleading vs. not misleading posts on social media that have been fact-checked by the crowd. Our key findings are as follows: (i) community fact-checked misleading tweets receive \SI{36.85}{\percent} fewer retweets than not misleading tweets (RQ1). (ii) There are significant differences in virality across different sub-types of misinformation (RQ2). Specifically, we find that misleading tweets are less viral than not misleading tweets across almost all sub-types of misinformation, except for (the relatively rare categories) satire, manipulated media, and outdated information. (iii) The fact-checking targets significantly differ between community fact-checkers and expert fact-checkers (RQ3). In particular, the crowd tends to fact-check posts from accounts with greater social influence (\eg, high-follower accounts). 

As an additional contribution, we conducted a user study to assess the perceived reliability of (real-world) community-created fact-checks (RQ4). We find that users agree with a relatively high share (\SI{73.33}{\percent}) of community-created fact-checks, whereas only a relatively small share (\SI{7.00}{\percent}) is perceived as being purposely deceptive (\eg, due to manipulation attempts). These results corroborate previous findings of experimental studies, which suggested that crowds can achieve a high level of accuracy when fact-checking social media content \cite[\eg,][]{Allen.2021}.

\textbf{Research implications:} In contrast to previous research examining the spread of misinformation that has been fact-checked by third-party organizations \cite{Vosoughi.2018,Solovev.2022b,Prollochs.2021b}, we find that community fact-checked misleading posts receive fewer retweets than not misleading posts. The diverging results may be a consequence of differences in the sample selection. 
While third-party organizations tend to fact-check posts on topics experts believe are of broad public interest and/or particularly concerning to society, community fact-checked posts comprise posts that have been deemed to be worth fact-checking by actual social media users. 

Our analysis suggests that crowd vs. experts focus on different targets when fact-checking social media content. We find that community fact-checkers tend to fact-check posts from larger accounts with high social influence, while expert fact-checks tend to target rumors shared by smaller accounts. Furthermore, community fact-checkers are relatively more likely to endorse/emphasize the accuracy of not misleading posts authored by influential users (\ie, users with a wide reach). This pattern is opposite to expert fact-checking where posts authored by influential accounts are relatively more likely to convey misinformation. Since author characteristics are inherently linked to the virality of posts (\eg, users with a wider reach can generate more retweets), the observed differences in fact-checking targets provide a (partial) explanation for the higher virality of not misleading community fact-checked posts. Note, however, that author characteristics are unlikely to be the only reason. In our explanatory regression analysis, we find the pattern that community fact-checked posts are more viral to persist -- even after controlling for the social influence of the author. This suggests that there might be additional differences between experts and the crowd in how fact-checking targets are selected (see \emph{Limitations and future research}). 

Importantly, while our study complements earlier work studying the diffusion of expert fact-checked posts, we do not claim that the selection by the crowd is more representative for the population of misinformation on social media as a whole. Rather, our results imply that the crowd focuses on different targets when fact-checking social media content and that sample selection plays a key role when studying misinformation diffusion. Compiling a representative sample of \emph{all} misinformation circulating on social media presents an important -- yet difficult -- challenge for future research.

\textbf{Practical implications:} From a practical perspective, policy initiatives around the world oblige social media platforms to develop countermeasures against misinformation. Community-based fact-checking opens new avenues to increase the scalability and speed of fact-checking of social media content. Furthermore, the community-based approach has the potential to to overcome trust issues associated with expert-created fact-checks \cite{Allen.2020}. 
The observed differences in the selection of fact-checking targets between community and expert fact-checkers suggest that both approaches might well complement each other. Here, community-created fact-checking may help to identify misinformation that is actually of interest to actual social media users -- and which may go unnoticed on third-party fact-checking organizations. 
The results of our user study further suggest that the vast majority of community-created fact-checks are perceived as being reliable. Although misuse of the platform cannot be prevented completely, previous research suggests that many issues with bad actors can effectively be addressed using sophisticated ranking mechanisms (\eg, helpfulness ratings), incentivizing high-quality fact-checks (\eg, blocking malicious contributors) or additional community-based content moderation efforts \cite{Epstein.2020,Godel.2021}. In sum, community-based fact-checking systems (as part of a larger toolset) allow social media platforms for improved coverage and may help to combat misinformation on social media more effectively. 

\textbf{Limitations and future research:} Our work has a number of limitations, which provide promising opportunities for future research. First, similar to related studies \cite[\eg,][]{Vosoughi.2018,Solovev.2022b}, we do not make causal claims. Future work should thus seek to validate our results in controlled experiments. Second, our user study evaluates the \emph{perceived} reliability of community-based fact-checks. While earlier experimental studies have already shown that crowds can achieve a high level of accuracy when fact-checking social media content \citeeg{Allen.2021}, it is necessary to further investigate the performance of the crowd in the field (\eg, via expert assessments of Birdwatch notes). Also, more research is necessary to better understand the role of manipulation attempts, and the conditions under which the wisdom of crowds can be unlocked for fact-checking. Third, our study shows that the fact-checking targets in community vs. expert fact-checks differ in terms of their author characteristics (\eg, number of followers). Future research should complement this analysis with a fine-grained study of additional characteristics of the fact-checked posts. For instance, it is a promising extension to employ topic modeling to study how the virality varies across topics (\eg, politics, health, entertainment, etc.) and other misinformation characteristics (\eg, novelty, believability). Fourth, our results are limited to Twitter's Birdwatch pilot. As such, the restricted set of Birdwatch contributors might not be representative for the overall user base on Twitter. Fifth, the community-created fact-checks in our study were not visible to the vast majority of Twitter users (\ie, only to pilot participants), whereas Twitter's goal is that Birdwatch will be available to everyone on Twitter. Future research may expand the current investigation by studying how (community-based) fact-checking \emph{labels} influence users' sharing behavior on social media. 

\section{Conclusion}

The spread of misinformation on social media is a pressing societal problem that platforms, policymakers, and researchers continue to grapple with. As a countermeasure, recent research proposed to build on crowd wisdom to fact-check social media content. In this study, we empirically analyzed the spread of posts that have been fact-checked by the crowd on Twitter's Birdwatch platform. Different from earlier studies that have analyzed the spread of misinformation fact-checked by third-party organizations, we find that crowd fact-checked misleading posts are less viral than not misleading posts. Our results also suggest that there are significant differences in virality across different sub-types of misinformation (\eg, factual errors, missing context, satire). Altogether, our findings offer insights into how misleading vs. not misleading posts spread and highlight the crucial role of sample selection when studying misinformation on social media. 

\section{Ethics Statement}
This research did not involve interventions with human subjects, and, thus, no approval from the Institutional Review Board was required by the authors' institutions. 


\bibliographystyle{ACM-Reference-Format-no-doi-abbrv}
\bibliography{literature}

\appendix

\newpage

\newpage
\vspace{1cm}
\begin{center}
	\huge Supplementary Materials
\end{center}
\vspace{1cm}

\section{Regression Results Without Outliers}

To assess the robustness of our analysis regarding outliers, we remove tweets with the top 1\% highest values for the retweet count. The results are presented Table \ref{table:robustness1}. All results are robust and confirm our previous findings.

\begin{table}[h]
\caption{Regression Results Without Outliers}
\label{table:robustness1}
\footnotesize
\begin{center}
\begin{tabular}{l S[table-format=2.3]S[table-format=2.3]S[table-format=2.3]}
\toprule
\multicolumn{4}{l}{Dependent Variable: Number of Retweets ($\mathit{RetweetCount})$}\\
\midrule
    & \multicolumn{1}{c}{{\begin{tabular}{@{}c@{}} \emph{Source Tweet}\\\end{tabular}}} 
		& \multicolumn{1}{c}{{\begin{tabular}{@{}c@{}} \emph{Fact-Checking Label}\\\end{tabular}}}
		& \multicolumn{1}{c}{{\begin{tabular}{@{}c@{}} \emph{Misinformation Types}\\\end{tabular}}}\\		
	\cmidrule(lr){2-2} \cmidrule(lr){3-3}  \cmidrule(lr){4-4}
 & \multicolumn{1}{c}{Model 1} & \multicolumn{1}{c}{Model 2} & \multicolumn{1}{c}{Model 3} \\
\midrule
Misleading                &               & -0.281^{***} &              \\
                          &              & (0.067)      &              \\

Factual Error             &              &              & -0.233^{***} \\
                          &              &              & (0.033)      \\
Missing Important Context &              &              & -0.093^{***} \\
                          &              &              & (0.032)      \\
Unverified Claim As Fact  &              &              & -0.273^{***} \\
                          &              &              & (0.032)      \\
Outdated Information      &              &              & 0.086^{*}    \\
                          &              &              & (0.052)      \\
Satire                    &              &              & 0.076        \\
                          &              &              & (0.075)      \\
Manipulated Media         &              &              & 0.479^{***}  \\
                          &              &              & (0.078)      \\
Other                     &              &              & -0.203^{***} \\
                          &              &              & (0.069)      \\
Delay                     & -0.018       & -0.017       & -0.028       \\
                          & (0.018)      & (0.018)      & (0.018)      \\
Sentiment                 & 0.076^{***}  & 0.074^{***}  & 0.070^{***}  \\
                          & (0.015)      & (0.015)      & (0.015)      \\
Followers                 & 0.214^{***}  & 0.212^{***}  & 0.213^{***}  \\
                          & (0.020)      & (0.020)      & (0.020)      \\
Followees                 & 0.133^{***}  & 0.134^{***}  & 0.144^{***}  \\
                          & (0.016)      & (0.016)      & (0.016)      \\
Account age               & -0.237^{***} & -0.238^{***} & -0.238^{***} \\
                          & (0.016)      & (0.016)      & (0.016)      \\
Verified                  & 1.063^{***}  & 1.070^{***}  & 1.114^{***}  \\
                          & (0.033)      & (0.033)      & (0.034)      \\
Intercept                 & 7.125^{***}  & 7.354^{***}  & 7.335^{***}  \\
                          & (0.087)      & (0.103)      & (0.090)      \\
Fixed effects (month-year)& {Yes}        & {Yes}        & {Yes}        \\
\midrule
AIC                       & \num{217196}   & \num{218106}   & \num{216960}   \\
Observations              & \num{15103}    & \num{15103}    & \num{15103}     \\
\bottomrule
\multicolumn{4}{r}{Significance levels: $^*p< 0.1$, $^{**}p< 0.05$, $^{***}p< 0.01$; standard errors in parentheses} \cr 
\multicolumn{4}{c}{\emph{Note:} Negative binomial regression explains the number of retweets of the fact-checked tweet.} \cr
\multicolumn{4}{c}{Month-year fixed effects are included.}
\end{tabular}
\end{center}

\end{table}

\newpage

\section{Separate Regressions for Misleading and Not Misleading Tweets}

We run separate for regressions for the subsets of misleading and not misleading tweets. The results remain robust (see Table~\ref{tbl:robustness2}). 

\begin{table}[h]
\caption{Regression Results for Subsets of Misleading and Not Misleading Tweets}
\label{tbl:robustness2}
\footnotesize
\begin{center}
\begin{tabular}{l S[table-format=2.3, table-column-width=4cm] S[table-format=2.3, table-column-width=3cm]}
\toprule
\multicolumn{3}{l}{Dependent Variable: Number of Retweets ($\mathit{RetweetCount})$}\\
\midrule
    & \multicolumn{1}{c}{{\begin{tabular}{@{}c@{}} Subset: \emph{Misleading}\\\end{tabular}}} 
		& \multicolumn{1}{c}{{\begin{tabular}{@{}c@{}} Subset: \emph{Not Misleading}\\\end{tabular}}}\\		
	\cmidrule(lr){2-2} \cmidrule(lr){3-3}  
	& \multicolumn{1}{c}{Model 1} & \multicolumn{1}{c}{Model 2} \\
\midrule
Delay          & -0.041^{**}  & -0.286^{***} \\
               & (0.019)      & (0.073)      \\
Sentiment      & 0.115^{***}  & -0.016       \\
               & (0.016)      & (0.070)      \\
Followers      & 0.273^{***}  & 0.270^{***}  \\
               & (0.020)      & (0.057)      \\
Followees      & 0.084^{***}  & 0.027        \\
               & (0.017)      & (0.047)      \\
Account age    & -0.224^{***} & -0.063       \\
               & (0.017)      & (0.077)      \\
Verified       & 0.803^{***}  & 0.504^{***}  \\
							 & (0.035)      & (0.158)      \\
Intercept      & 7.751^{***}  & 8.416^{***}  \\
               & (0.098)      & (0.234)      \\
Fixed effects (month-year)& {Yes}        & {Yes}    \\
\midrule
AIC            & \num{209875}   & \num{13287}   \\
Observations   & \num{14384}    & \num{872}         \\
\bottomrule
\multicolumn{3}{r}{Significance levels: $^*p< 0.1$, $^{**}p< 0.05$, $^{***}p< 0.01$; standard errors in parentheses} \cr 
\multicolumn{3}{c}{\emph{Note:} Negative binomial regression explains the number of retweets of the fact-checked tweet.} \cr
\multicolumn{3}{c}{Month-year fixed effects are included.}

\end{tabular}
\end{center}
\end{table}

\newpage
\section{Variance Inflation Factors}

We calculated variance inflation factors for all explanatory variables in our regression models for RQ1 and RQ2 (Table~\ref{tbl:results_vif}). The VIFs are substantially below the critical threshold of four. This indicates that multicollinearity is not an issue in our analysis.

\begin{table}[H]
\caption{Variance Inflation Factors for Regression Models}
\label{tbl:results_vif}
{
\sisetup{table-space-text-post = {}} 
\footnotesize
\begin{tabular}{lS[table-format=1.3,round-mode=places,round-precision=3]S[table-format=1.3,round-mode=places,round-precision=3]}
\toprule
  & \mc{\textbf{RQ1}} & \mc{\textbf{RQ2}}   \\
\midrule
Misleading                & 1.018925 &    \\
Delay                     & 1.004293 &  1.007520      \\
Sentiment                 & 1.010687 & 1.016269    \\
Followers                 & 1.069464 & 1.072113    \\
Followees                 & 1.001402 & 1.002832    \\
Account Age               & 1.155483 & 1.176560    \\
Verified                  & 1.193135 & 1.251476    \\
Factual Error             &          &  1.092877 \\
Missing Important Context &          &  1.082110  \\
Unverified Claim As Fact  &          &  1.129219  \\
Outdated Information      &          &  1.045445 \\
Satire                    &          &  1.048804 \\
Manipulated Media         &          &  1.052999  \\
Other                     &          &  1.020949  \\
\bottomrule
\end{tabular}
}
\end{table}

\newpage
\section{Analysis With User-Specific Random Effects}

Fact-checks on Birdwatch are performed by many different contributors. To account for this, we include random effects for the individual Birdwatch contributors into our regression model. The regression results are reported in Table \ref{tbl:robustness3}. All results are robust and confirm our previous findings.

\begin{table}[h]
\caption{Regression Results With User-Specific Random Effects}
\label{tbl:robustness3}
\footnotesize
\begin{center}
\begin{tabular}{l S[table-format=2.3]S[table-format=2.3]S[table-format=2.3]}
\toprule
\multicolumn{4}{l}{Dependent Variable: Number of Retweets ($\mathit{RetweetCount})$}\\
\midrule
    & \multicolumn{1}{c}{{\begin{tabular}{@{}c@{}} \emph{Source Tweet}\\\end{tabular}}} 
		& \multicolumn{1}{c}{{\begin{tabular}{@{}c@{}} \emph{Fact-Checking Label}\\\end{tabular}}}
		& \multicolumn{1}{c}{{\begin{tabular}{@{}c@{}} \emph{Misinformation Types}\\\end{tabular}}}\\		
	\cmidrule(lr){2-2} \cmidrule(lr){3-3}  \cmidrule(lr){4-4}
 & \multicolumn{1}{c}{Model 1} & \multicolumn{1}{c}{Model 2} & \multicolumn{1}{c}{Model 3} \\
\midrule
Misleading                &              & -0.456^{***} &              \\
                          &              & (0.068)      &              \\

Factual Error             &              &              & -0.251^{***} \\
                          &              &              & (0.034)      \\
Missing Important Context &              &              & -0.127^{***} \\
                          &              &              & (0.033)      \\
Unverified Claim As Fact  &              &              & -0.300^{***} \\
                          &              &              & (0.033)      \\
Outdated Information      &              &              & -0.061       \\
                          &              &              & (0.054)      \\
Satire                    &              &              & 0.411^{***}  \\
                          &              &              & (0.077)      \\
Manipulated Media         &              &              & 0.462^{***}  \\
                          &              &              & (0.080)      \\
Other                     &              &              & -0.222^{***} \\
                          &              &              & (0.071)      \\
Delay                     &              & -0.048^{***} & -0.054^{***} \\
                          &              & (0.018)      & (0.018)      \\				
Sentiment                 & 0.068^{***}  & 0.068^{***}  & 0.061^{***}  \\
                          & (0.014)      & (0.014)      & (0.014)      \\							
Followers                 & 0.271^{***}  & 0.267^{***}  & 0.268^{***}  \\
                          & (0.019)      & (0.019)      & (0.019)      \\
Followees                 & 0.074^{***}  & 0.076^{***}  & 0.086^{***}  \\
                          & (0.016)      & (0.016)      & (0.016)      \\
Account age               & -0.213^{***} & -0.216^{***} & -0.227^{***} \\
                          & (0.017)      & (0.017)      & (0.017)      \\
Verified                  & 0.774^{***}  & 0.783^{***}  & 0.866^{***}  \\
                          & (0.034)      & (0.034)      & (0.035)      \\
Intercept                 & 7.922^{***}  & 8.266^{***}  & 8.122^{***}  \\
                          & (0.089)      & (0.105)      & (0.092)      \\
Fixed effects (month-year)& {Yes}        & {Yes}        & {Yes}        \\
Random effects (user)     & {Yes}        & {Yes}        & {Yes}        \\
\midrule
AIC                       & \num{223233}   & \num{223182}   & \num{222893}   \\
Observations              & \num{15256}    & \num{15256}    & \num{15256}     \\
\bottomrule
\multicolumn{4}{r}{Significance levels: $^*p< 0.1$, $^{**}p< 0.05$, $^{***}p< 0.01$; standard errors in parentheses} \cr 
\multicolumn{4}{c}{\emph{Note:} Negative binomial regression explains the number of retweets of the fact-checked tweet.} \cr
\multicolumn{4}{c}{Month-year fixed effects are included.}
\end{tabular}
\end{center}
\end{table}

\newpage
\section{Quadratic Effects and Interaction Terms}

As a robustness check, we include quadratic effects and interaction terms between the fact-checking label and the source tweet variables into our regression analysis. The results remain robust and support our findings (see Table~\ref{tbl:robustness4}).

\begin{table}[h]
\caption{Regression Results With Quadratic Effects and Interaction Terms}
\label{tbl:robustness4}
\scriptsize
\begin{center}
\begin{tabular}{l S[table-format=2.3, table-column-width=4cm] S[table-format=2.3, table-column-width=3cm]}
\toprule
\multicolumn{3}{l}{Dependent Variable: Number of Retweets ($\mathit{RetweetCount})$}\\
\midrule
    & \multicolumn{1}{c}{{\begin{tabular}{@{}c@{}} \emph{Quadratic Effects}\\\end{tabular}}} 
		& \multicolumn{1}{c}{{\begin{tabular}{@{}c@{}} \emph{Interaction Terms}\\\end{tabular}}}\\		
	\cmidrule(lr){2-2} \cmidrule(lr){3-3}  
	& \multicolumn{1}{c}{Model 1} & \multicolumn{1}{c}{Model 2} \\
\midrule

Misleading                      & -0.479^{***} & -0.659^{***} \\
                                & (0.068)      & (0.106)      \\
Delay                           & 0.086^{**}   & -0.242^{***} \\
                                & (0.042)      & (0.070)      \\
Delay$^2$                       & -0.010^{***} &              \\
                                & (0.003)      &              \\
Sentiment                       & 0.136^{***}  & 0.020        \\
                                & (0.016)      & (0.066)      \\
Sentiment$^2$                   & 0.011        &              \\
                                & (0.007)      &              \\
Followers                       & 0.548^{***}  & 0.249^{***}  \\
                                & (0.045)      & (0.054)      \\
Followers$^2$                   & -0.029^{***} &              \\
                                & (0.006)      &              \\
Followees                       & 0.115^{***}  & 0.025        \\
                                & (0.024)      & (0.045)      \\
Followees$^2$                   & -0.003^{*}   &              \\
                                & (0.002)      &              \\
Account age                     & -0.416^{***} & -0.058       \\
                                & (0.022)      & (0.073)      \\
Account age$^2$                 & -0.341^{***} &              \\
                                & (0.021)      &              \\
Verified                        & 0.739^{***}  & 0.389^{***}  \\
                                & (0.035)      & (0.150)      \\
Misleading $\times$ Delay       &              & 0.201^{***}  \\
                                &              & (0.072)      \\
Misleading $\times$ Sentiment   &              & 0.096        \\
                                &              & (0.068)      \\
Misleading $\times$ Followers   &              & 0.024        \\
                                &              & (0.058)      \\
Misleading $\times$ Followees   &              & 0.059        \\
                                &              & (0.048)      \\
Misleading $\times$ Account age &              & -0.166^{**}  \\
                                &              & (0.075)      \\
Misleading $\times$ Verified    &              & 0.416^{***}  \\
                                &              & (0.155)      \\
Intercept                       & 8.641^{***}  & 8.436^{***}  \\
                                & (0.107)      & (0.130)      \\
Fixed effects (month-year)      & {Yes}        & {Yes}    \\
\midrule
AIC            & \num{222906}   & \num{223179}   \\
Observations   & \num{15256}        & \num{15256}        \\
\bottomrule
\multicolumn{3}{r}{Significance levels: $^*p< 0.1$, $^{**}p< 0.05$, $^{***}p< 0.01$; standard errors in parentheses} \cr 
\multicolumn{3}{c}{\emph{Note:} Negative binomial regression explains the number of retweets of the fact-checked tweet.} \cr
\multicolumn{3}{c}{Month-year fixed effects are included.}
\end{tabular}
\end{center}
\end{table}

\newpage
\section{Alternative Handling of Multiple Fact-Checks}
\label{appendix:Consensus}

Our main analysis focuses on the temporally first fact-check after the tweet has been posted. As a robustness check, we evaluate whether our results are robust to alternative handling of multiple fact-checks. We repeated our analysis with the following variants: (i) we determined the fact-checking label via majority vote; (ii) we use Birdwatch's rating mechanism (see \cite{Twitter.2021} for details) to identify the fact-check with which most users agree; (iii) we consider all fact-checks without any filtering. 

The regression results are presented in Table \ref{table:robustness_multiple}. In all cases, we find qualitatively identical results that support our previous findings.

\begin{table}[h]
\caption{Regression Results With Alternative Handling of Multiple Fact-Checks}
\label{table:robustness_multiple}
\footnotesize
\begin{center}
\begin{tabular}{l S[table-format=2.3, table-column-width=1.3cm]S[table-format=2.3, table-column-width=1.3cm]S[table-format=2.3, table-column-width=1.3cm]S[table-format=2.3, table-column-width=1.3cm]S[table-format=2.3, table-column-width=1.3cm]S[table-format=2.3, table-column-width=1.3cm]}
\toprule
\multicolumn{7}{l}{Dependent Variable: Number of Retweets ($\mathit{RetweetCount})$}\\
\midrule
    & \multicolumn{2}{c}{{\begin{tabular}{@{}c@{}} \emph{(i) Majority Vote}\\\end{tabular}}} 
		& \multicolumn{2}{c}{{\begin{tabular}{@{}c@{}} \emph{(ii) Highest Agreement}\\\end{tabular}}}
		& \multicolumn{2}{c}{{\begin{tabular}{@{}c@{}} \emph{(iii) All Fact-Checks}\\\end{tabular}}} \\		
	\cmidrule(lr){2-3} \cmidrule(lr){4-5}  \cmidrule(lr){6-7}
 & \multicolumn{1}{c}{Model 1} & \multicolumn{1}{c}{Model 2} & \multicolumn{1}{c}{Model 3} & \multicolumn{1}{c}{Model 4} & \multicolumn{1}{c}{Model 5} & \multicolumn{1}{c}{Model 6} \\
\midrule
Misleading          &              &  -0.630^{***}&              & -0.728^{***} &              & -0.849^{***} \\
                    &              &  (0.068)     &              & (0.057)      &              & (0.043)      \\
Delay               &              & 0.018        &              & 0.011        &              & -0.004       \\
                    &              & (0.017)      &              & (0.018)      &              & (0.013)      \\
Sentiment           & 0.115^{***}  & 0.112^{***}  & 0.112^{***}  & 0.109^{***}  & 0.046^{***}  & 0.064^{***}  \\
                    & (0.016)      & (0.016)      & (0.016)      & (0.016)      & (0.013)      & (0.013)      \\
Followees           & 0.087^{***}  & 0.092^{***}  & 0.074^{***}  & 0.078^{***}  & 0.021        & 0.033^{**}   \\
                    & (0.017)      & (0.017)      & (0.016)      & (0.016)      & (0.013)      & (0.013)      \\
Followers           & 0.284^{***}  & 0.273^{***}  & 0.271^{***}  & 0.257^{***}  & 0.319^{***}  & 0.316^{***}  \\
                    & (0.020)      & (0.020)      & (0.019)      & (0.019)      & (0.014)      & (0.014)      \\
Account age         & -0.223^{***} & -0.223^{***} & -0.213^{***} & -0.214^{***} & -0.190^{***} & -0.192^{***} \\
                    & (0.017)      & (0.017)      & (0.017)      & (0.017)      & (0.014)      & (0.014)      \\
Verified            & 0.789^{***}  & 0.796^{***}  & 0.774^{***}  & 0.763^{***}  & 0.717^{***}  & 0.731^{***}  \\
                    & (0.035)      & (0.035)      & (0.034)      & (0.034)      & (0.029)      & (0.029)      \\
(Intercept)         & 7.879^{***}  & 8.375^{***}  & 7.922^{***}  & 8.477^{***}  & 8.821^{***}  & 9.559^{***}  \\
                    & (0.091)      & (0.107)      & (0.089)      & (0.100)      & (0.070)      & (0.079)      \\
Fixed effects (month-year) & {Yes} & {Yes}        & {Yes}        & {Yes}        & {Yes}        & {Yes}       \\
\midrule
AIC                 & \num{211311} & \num{211213} & \num{223233} & \num{223039} & \num{317871} & \num{317409}   \\
Observations        & \num{14619}  & \num{14619}  & \num{15256}  & \num{15256}  & \num{20218}  & \num{20218}        \\
\bottomrule
\multicolumn{7}{r}{Significance levels: $^*p< 0.1$, $^{**}p< 0.05$, $^{***}p< 0.01$; standard errors in parentheses} \cr 
\multicolumn{7}{c}{\emph{Note:} Negative binomial regression explains the number of retweets of the fact-checked tweet.} \cr
\multicolumn{7}{c}{Month-year fixed effects are included.}
\end{tabular}
\end{center}
\end{table}

%
%

\end{document}